\documentclass[10pt]{iopart}

\usepackage{cite}
\usepackage{booktabs}
\usepackage{standalone}
\usepackage{tikz,pgfplots}
\usetikzlibrary{spy,backgrounds}
\usepackage{comment}
\usepackage{appendix}
\usepackage[breaklinks=true,
            colorlinks=true,
            linkcolor=blue,
            urlcolor=blue,
            citecolor=blue,
            bookmarksopen=true,
            bookmarksopenlevel=2,
            bookmarksnumbered=true,
            pdfauthor=Maximilian Dreisbach,
            pdfstartview=FitH]
            {hyperref}
\usepackage{xcolor}
\usepackage{amssymb}
\usepackage{caption}
\usepackage{subfigure}
\pgfplotsset{compat=1.9}
\usepackage{textgreek}

\definecolor{KITred}{RGB}{162, 34, 35}
\definecolor{KITblue}{RGB}{70 100 170}
\definecolor{IntM}{rgb}{0.00000,0.44700,0.74100}

\newcommand{\eqref}[1]{(\ref{#1})}
\newcommand{\citet}[1]{\emph{et al.}\,\cite{#1}}
\begin{document}

\title[Spatio-temporal reconstruction of drop impact dynamics by means of color-coded glare points and deep learning]{Spatio-temporal reconstruction of droplet impingement dynamics by means of color-coded glare points and deep learning}

\author{Maximilian Dreisbach$^{1}$, Jochen Kriegseis$^1$, Alexander Stroh$^{1}$}

\address{$^1$ Institute of Fluids Mechanics (ISTM), Karlsruhe Institute of Technology (KIT), Kaiserstraße 10, 76131 Karlsruhe, Germany}

\vspace{10pt}
\begin{indented}
\item[]September 2023
\end{indented}

\begin{abstract}

In order to capture the complex three-dimensional shape of the gas-liquid interface in two-phase flows commonly multiple viewpoints and elaborate reconstruction methods are required, in particular for highly deformed interfaces that result in self-occlusion.
The present work introduces a deep learning approach for the three-dimensional reconstruction of the spatio-temporal dynamics of the gas-liquid interface on the basis of monocular images obtained via optical measurement techniques.
The method is tested an evaluated at the example of liquid droplets impacting on structured solid substrates.
The droplet dynamics are captured through high-speed imaging in an extended shadowgraphy setup with additional reflective glare points from lateral light sources that encode further three-dimensional information of the gas-liquid interface in the images.
A neural network is learned for the physically correct reconstruction of the droplet dynamics on a labelled dataset generated by synthetic image rendering on the basis of gas-liquid interface shapes obtained from direct numerical simulation.
The employment of synthetic image rendering allows for the efficient generation of training data and circumvents the introduction of errors resulting from the inherent discrepancy of the droplet shapes between experiment and simulation. 
The accurate reconstruction of the three-dimensional shape of the gas-liquid interface during droplet impingement on the basis of images obtained in the experiment demonstrates the practicality of the presented approach based on neural networks and synthetic training data generation.
The introduction of glare points from lateral light sources in the experiments was shown to improve the reconstruction accuracy, which indicates that the neural network learns to leverage the additional three-dimensional information encoded in the images for a more accurate depth estimation.
By the successful reconstruction of obscured areas in the input images, it is demonstrated that the neural network has the capability to learn a physically correct interpolation of missing data from the numerical simulation.
Furthermore, the physically reasonable reconstruction of unknown gas-liquid interface shapes for drop impact regimes that were not contained in the training dataset indicates that the neural network learned a versatile model of the involved two-phase flow phenomena during droplet impingement.

\end{abstract}

\ioptwocol


\section{Introduction}
\label{sec:introduction}
Droplet impingement on wet or dry surfaces is a relevant phenomenon in a multitude of technical applications, such as spray cooling \cite{Ashgriz2011}, spray coating \cite{Andrade2013,Dalili2016}, inkjet printing \cite{Lohse2022} and combustion \cite{Moreira2010}.
For example, in spray cooling an optimisation of the droplet-wall interaction leads to an increase in efficiency \cite{Ashgriz2011}, whereas in spray coating the prevention of air bubble entrapment necessary to ensure a high quality of the surface \cite{Dalili2016}.
The outcome of droplet impingement is dependant on kinematic conditions, in particular the impact velocity, the angle of incidence and the volume of the droplet, as well as fluid properties, most importantly the surface tension, viscosity and density of the liquid droplet.
If the impact on a solid substrate is considered, the structure and roughness of the substrate influence whether a deposition, splashing, a partial or complete rebound of the droplet \cite{Rioboo2001} occurs.
Due to variety of outcomes and their complex dependence on the impact conditions, various details of these droplet impingement phenomena remain to be understood \cite{Josserand2016}.
Droplet impingement is an inherently three-dimensional (3D) process due to the heterogeneity of the substrate that can be chemical or roughness and outside perturbation from the environment.
In particular, angled impacts and the anisotropic wetting of structured surfaces lead to a complicated three-dimensional deformation of the gas-liquid interface.

The most common optical measurement method for the observation of droplet dynamics, due to its simplicity and high spatial accuracy, is the shadowgraphy technique, in which a droplet or bubble is illuminated in parallel backlight that accurately maps the contour of the gas-liquid interface onto an image plane \cite{Tropea2007,Nitsche2006}.

In order to capture the complex three-dimensional deformation of the droplet during impingement on structured surfaces, commonly multiple camera angles are required.
Numerical simulation can accurately predict the three-dimensional shape of the gas-liquid interface \cite{fink2018,worner2021}, however their model functions for the contact angle rely on experimental measurements \cite{kistler1993,cox1986}.
Furthermore, experimental data is required for validation.
Therefore, the extraction of three-dimensional data from the experiments would allow for a deeper insight into the dynamics of the droplet during the impingement on structured surfaces.
Various experimental techniques for the volumetric reconstruction of droplets and bubbles have been developed.
More recently, advances in deep learning techniques have lead to impressive results for the data-driven reconstruction of complex three-dimensional shapes from multiple or even a single viewpoint.

\subsection{Volumetric reconstruction of gas-liquid interfaces}
\label{subsec:gas_liquid_interface_reconstruction}

For the reconstruction of the three-dimensional gas-liquid interface in two-phase flows various measurement techniques based on different optical phenomena have been proposed, including methods based on refraction \cite{Morris2011,Qian2017,Dehaeck2013,Dehaeck2015}, fluorescence \cite{Ihrke2005, Roth2020}, light scattering \cite{Konig1986,Glover1995,Dehaeck2005,Brunel2021,Dreisbach2023} and structured light techniques \cite{Horbach2010,Zhang2015,Hu2015}.
The most common approach for the volumetric reconstruction of gas-liquid interfaces in bubbles or droplets is based on multi-view shadowgraphy experiments.
Tomiyama \citet{Tomiyama2002} determine the integral volume of a bubble from a single shadowgraph image. The authors assume an oblate spheroidal shape of the bubble and determine its major and minor axis from the shadowgraph contour.
Fujiwara \citet{Fujiwara2004} reconstruct the 3D-shape of deformed bubbles from two orthogonal shadowgraph views by fitting multiple cross sections along the third orthogonal direction as ellipses whose main axis are estimated from the horizontal extent of the bubble in the two views.
Honkanen \citet{Honkanen2009} extend this slicing approach by first determining an oriented 3D bounding box for the bubble, in which then horizontal cross sections are fitted by ellipses.
Fu and Liu \cite{Fu2018} employ the space carving technique \cite{Laurentini1994} to combine the contours of a bubble at four different viewing angles into a virtual hull, which represents a maximum estimate of the bubble volume. 
Subsequently this virtual hull is smoothed by spline fits on multiple cross sections in a slicing approach, in order to represent the effects of surface tension and reach a more accurate reconstruction of the three-dimensional bubble shape.
Masuk \citet{Masuk2019} introduce additional virtual cameras to the space carving technique in order to consider surface tension.
In their approach the reconstructed hull is projected to the novel views of virtual cameras, in which locations of high curvature are iteratively smoothed, while the contour of the gas-liquid interface in all real views is respected.
R{\'i}os-L{\'o}pez \citet{RiosLopez2018} reconstruct the 3D-shape of a deformed, non-axisymmetrical droplet sliding on a flat surface from two orthogonal views through a polynomial fit of the contours with the assumption of plane symmetry.
More recently Gong \citet{Gong2022} learned a neural network for the volumetric reconstruction of one side of a bubble in form of a depth map from grayscale information of a single shadowgraph image. The authors employed a pyramidal convolutional neural network \citet{Lecun2015} that was trained on rendered synthetic images of bubbles and their respective ground truth volumetric shapes.

For the measurement of flow properties in two phase flows, e.g. the diameter and position of bubbles, various methods based on glare points from light scattering on the gas-liquid interface have been developed.
Glare points originate from interface reflection or refraction with subsequent transmission and can partake in an arbitrary number of internal reflections within a droplet or bubble, before the light ray exits the gas-liquid interface again, casting a glare point at the exit location of the rays.
Most methods are based on the interference patterns created by defocused glare points \cite{Konig1986,Glover1995,Brunel2021}, but with the availability of high resolution camera sensors methods based on in-focus glare points have become feasible as well.
Dehaeck \citet{Dehaeck2005} demonstrate how the relative position of glare points from two light sources can be used to obtain the aspect ratio and the tilt angle of a non-spherical bubbles, as well as the relative refractive index of the two involved fluids.
Dreisbach \citet{Dreisbach2023} employ glare points to encode the complex 3D-shape of a deforming droplet during impingement and show that the aspect ratio and observation angle of a droplet in a non-isotropic wetting state can be reconstructed from the relative glare point distances.

\subsection{Deep learning for volumetric representations}
\label{subsec:deep_learning_rec}

In recent years, deep learning methods the for volumetric reconstruction from images have evolved rapidly and are a promising prospect for the reconstruction of multi-phase flows.
The principal idea is learning a neural network for the representation of a deformable 3D-geometry on a large paired dataset of input images and output 3D-shapes or even just multi-view images.
The prior knowledge learned by the neural network resolves ambiguities in the input and thereby allows for the volumetric reconstruction of the 3D-geometry from as little as a single image.
Different 3D-representations have been proposed in contemporary works, namely voxel-based \cite{Girdhar2016, Choy2016,Wu2016,Riegler2017}, point cloud \cite{Fan2017, Lin2018} and mesh-based \cite{Wang2018}, implicit representations \cite{Chen2019, Park2019, Mescheder2019, saito2019} and neural rendering techniques \cite{Niemeyer2020, Mildenhall2020}.

Early works lean on the success of convolutional neural networks (CNN) \cite{Lecun2015} as the 2D convolution operation can straight-forwardly be extended to the 3D-domain for the prediction of discrete volumetric representations, such as voxel grids.
Girdhar \citet{Girdhar2016} learn a joint low dimensional representation for 2D-images and 3D voxel shapes through a 3D convolutional auto-encoder network \cite{Kingma2013}, which allows for the reconstruction of 3D-geometries from a single image.
Choy \citet{Choy2016} employ a recurrent long short-term memory network (LSTM) \cite{Hochreiter1997} that consecutively processes multiple images and efficiently merges information of previously unseen parts of the 3D-geometry with each novel viewpoint.
Wu \citet{Wu2016} combine generative adversarial training \cite{Goodfellow2020} with 3D convolutional neural networks, which lead to a more realistic shape generation.
Furthermore, through the integration of variational autoencoders (VAE) \cite{Kingma2013} their 3D-VAE-GAN architecture allows for the volumetric reconstruction from a single image.
While voxel representations can handle arbitrary topology and deliver accurate results, they are not suitable for the reconstruction of fine details, as the computational and memory requirements grow cubically with the resolution.
Furthermore, voxels are an inefficient representation of a 3D-geometry, since information lies only on the surface voxels.
Riegler \citet{Riegler2017} propose a voxel representation through adaptive hierarchical octary trees \cite{Meagher1982}, which have a fine resolution near the surface and a coarse resolution in the rest of the domain, thus alleviating the drawbacks of voxel representations. 

The representation of a 3D-geometry as a mesh or point cloud allows for a more compact and scalable encoding of the surface, with low memory and computational requirements.
Fan \citet{Fan2017} propose to learn a point cloud representation of the 3D-geometry for the volumetric reconstruction from monocular images.
To that regard the authors introduce a novel conditional generative network, consisting of a CNN-based image encoder, followed by two branching paths for point cloud prediction, a deconvolution network that preserves spatial continuity and a multi-layer perceptron (MLP) that predicts accurate fine details. 
Lin \citet{Lin2018} employ a 2D-convolutional encoder-decoder network to predict multi-channel images at different novel viewpoints as an intermediate representation, which encode the coordinates of a point cloud.
Subsequently, these intermediate representations are fused into a dense point cloud of the 3D-surface through a transformation into a canonical space
Point cloud representation are simple to implement and to learn with neural networks, since they are unordered and no connectivity needs to be represented, however they require considerable effort in post-processing for the retrieval of the 3D-geometry.
Wang \citet{Wang2018} learn a mesh-based representation through a graph based fully convolutional network, that reconstructs a 3D-geometry by deformation of a template mesh in a coarse-to-fine manner.
Their mesh-based representation allows for information flow between neighbouring vertices during training, which helps to provide a regular smooth output, however the topology is restricted by the mesh template.

More recently implicit representations of continuous 3D-shapes through level set functions that are approximated through neural networks have been proposed.
Chen \citet{Chen2019} and Mescheder \citet{Mescheder2019} learn a MLP for the implicit field representation of a 3D-shape as an occupancy function, that takes the value of one if a point coordinate lies inside of the shape and zero otherwise.
This approach be seen as effectively learning a binary classification network for the approximation of a decision boundary that represents the surface.
Park \citet{Park2019} propose the representation by a signed distance function (SDF) that takes values greater zero on the outside and values smaller than zero on the inside, thus placing the surface at $SDF=0$.
A MLP evaluates this implicit function for randomly sampled 3D-coordinates that are concatenated with the global image features extracted from an input image, thus making the method suitable for monocular volumetric reconstruction.
Due to the continuous nature of the implicit representation it can be evaluated at any arbitrary resolution and consequently a high surface quality can be reached, while the memory requirement is comparably low.
The marching cubes algorithm \cite{Lorensen1987} is commonly employed to  reconstruct a coherent 3D-mesh from the evaluated point coordinates.
Since methods based on implicit functions rely on global context for the prediction of 3D-shapes, the local alignment with the input image is not guarantied. 
Saito \citet{saito2019} propose the extraction of pixel-aligned local features by first processing the input image through a fully convolutional hourglass network \cite{Newell2016} prior to the prediction of a level set function with a MLP.
The combination of local features with a global 3D representation by an implicit function allows for the reconstruction of fine details, while the global shape is preserved.

Recently, differentiable rendering techniques have become increasingly relevant, as they show impressive results in novel view synthesis and volumetric reconstruction that can be learned without 3D-supervision.
Niemeyer \citet{Niemeyer2020} employ differentiable volumetric rendering and an implicit representation of the 3D-geometry for monocular volume and texture reconstruction.
Mildenhall \citet{Mildenhall2020} introduced neural radiance fields, a continuous volumetric representation of a scene that is encoded inside a simple multi-layer perceptron.
From this representation novel views from any arbitrary angle can be obtained by volumetric rendering.

Currently, the volumetric reconstruction of the deformed gas-liquid interfaces in two-phase flows from shadowgraphy experiments typically requires multiple viewpoints, in particular for cases in which a large deformation causes self-occlusion, such as during droplet impingement.
However, the use of a single view technique would allow for a simple set up and calibration, as well as an affordable experimental apparatus.
Furthermore, limited optical access to the flow could restrict the experiments to single-view techniques.
The present work, therefore, proposes the employment of data-driven techniques for the spatio-temporal reconstruction of the 3D gas-liquid interface of droplets during impingement from experiments with a single viewpoint.
To that regard the canonical shadowgraphy technique is modified by color-coded glare points from additional lateral light sources that encode further information on the shape of the gas-liquid interface, which serves as the basis for the reconstruction by a neural network.

\section{Methodology}
\label{sec:methodology}

The present work introduces a deep learning approach for the volumetric reconstruction of the gas-liquid interface during droplet impingement from monocular images obtained through an extended shadowgraphy technique. 
A neural network is trained for the physically correct reconstruction by supervised learning on a large dataset of labelled data, i.e pairs of matching input images and respective output three-dimensional shapes.
From numerical simulation three-dimensional ground truth data for the gas-liquid interface is available, whereas the input images are obtained through optical measurement techniques.
However, due to uncertainty in the experiments and errors from modelling and approximations in the simulation the droplet shapes are not identical, resulting in a mismatch of the input images with their respective ground truth labels that consequently introduces an error to the neural network.
This kind of matching problem is resolved by training the neural network on fully synthetic image data that is obtained using the ground truth of the simulation, thus results in a perfect agreement of input and output in the dataset.
The synthetic images are generated by means of a render-pipeline in \textit{Blender} with the \textit{LuxCore} package that allows for physically correct ray-tracing.
The virtual rendering setup mirrors the optical configuration in the experiments in order to produce realistic image data.
The PIFu neural network \cite{saito2019} is learned for the spatio-temporal reconstruction of the droplet dynamics during impingement on this synthetic dataset.
First, the experimental setup is presented in subsection~\ref{subsec:experimental_setup}, followed by the explanation of the synthetic data acquisition in subsection~\ref{subsec:synthetic_data} and finally the network training in subsection~\ref{subsec:network_training}.
Afterwards, the methodology for the evaluation of the reconstruction result is presented in subsection~\ref{subsec:eval_metrics}.

\subsection{Experimental setup}
\label{subsec:experimental_setup}

\begin{figure}[htbp!]
    \centering
    \includegraphics[width=1.0\columnwidth]{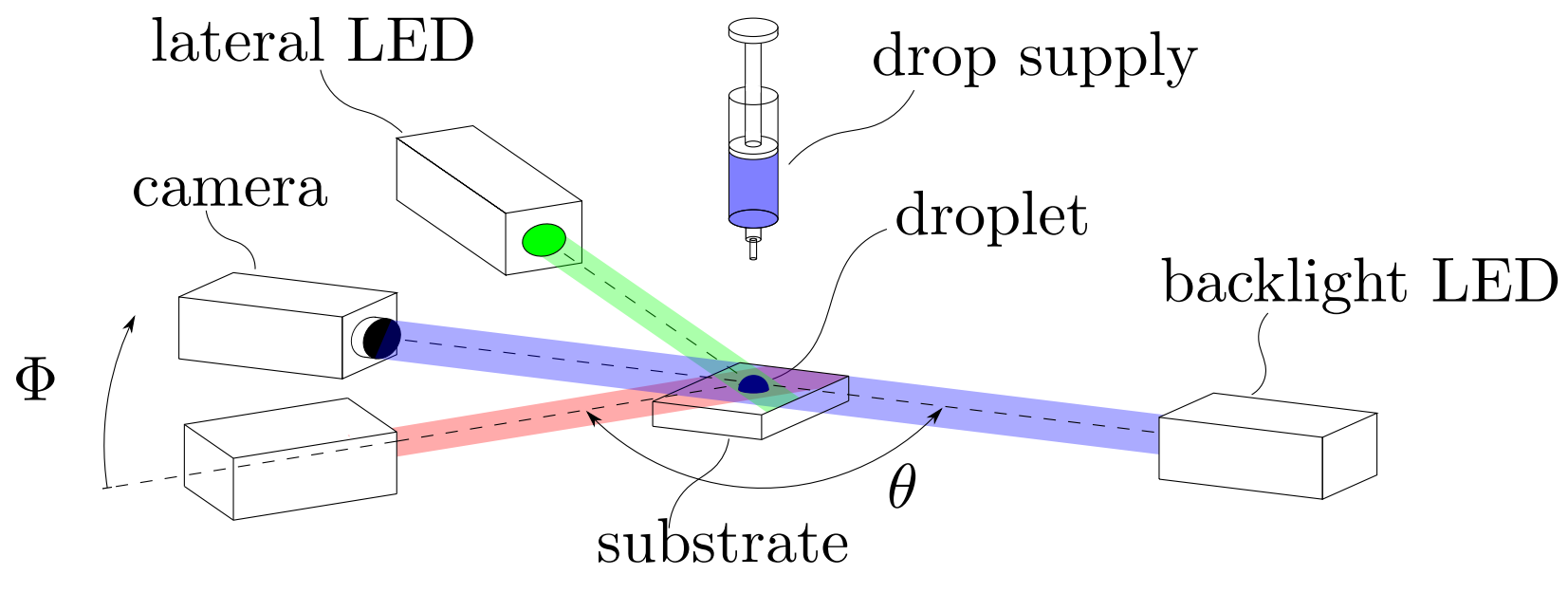}
    \caption{Sketch of the measurement setup, indicated are the scattering angle $\theta$ and the elevation angle $\Phi$. Figure adopted from \cite{Dreisbach2023}.}
    \label{fig:exp_setup}
\end{figure}

In order to facilitate the volumetric reconstruction from a monocular video, an experimental method that represents the three-dimensional shape of the deforming gas-liquid interface in the two-dimensional images is employed.
The basis of the imaging setup is the standard shadowgraphy technique, which allows for the observation of an accurate contour of the gas-liquid interface.
In a previous work the authors introduced colored glare points from lateral light sources to the canonical setup in order to encode additional three-dimensional information in the image \cite{Dreisbach2023}.
A sketch of the experimental setup can be seen in Figure~\ref{fig:exp_setup} and the resulting images in Figure~\ref{fig:comp_render_real}~(a).

The lateral illumination is produced by high-power \emph{ILA\_5150 LPSv3} LEDs with narrow-banded spectra and maxima in the visible spectrum at $\sim 455$\,nm ("\emph{blue}"), $ \sim 521$\,nm ("\emph{green}") and $\sim 632$\,nm ("\emph{red}").
In order to ensure a reproducible volume of the droplet an automated drop supply is used, consisting of a syringe with a cannula diameter of $d_s = 0.1$\,mm that is actuated by a linear stepper motor.
Three solid substrates are considered, a flat silicon oxide (SiOx) surface, a structured Polydimethylsiloxane (PDMS) substrate with regular $60$\,$\mu m$ square grooves and a structured 3D-printed polylactide (PLA) substrate with a spacing of $154 \mu$m.
The impinging droplet is imaged by a \emph{Photron Nova R2} equipped with a \emph{Schneider-Kreuznach Apo-Componon} $4.0/60$ enlarging lens at $7,500$ frames per second (fps) and $1,280$\,px x $512$\,px resolution.

The glare points from the lateral light sources result from specular reflection on the gas-liquid interface of the droplet, whereas the blue glare point from the backlight is produced by transmission through the droplet (see Figure~\ref{fig:masking}).
The location and intensity of the glare points is dependent on the scattering angle $\theta$ and the order of the glare points $p$, which is defined by the number of chords travelled through the droplet \cite{Hulst1981}.
Consequently, the specular glare points from the lateral light sources are referred as $p=0$ glare points and the glare point from the backlight as $p=1$ in the following.
It should be mentioned that in general glare points of orders $p\geq1$ can be visible for different scattering angles.
However, for simplicity a scattering angle of $\theta = 95.6^\circ$ was set for the lateral light sources, as it excludes higher order glare points ($p\geq1$).
Furthermore, the elevation angle of the lateral light sources was set to $\Phi=45^\circ$ in order to ensure the emergence of glare points in all frames.

For a given experimental setup in which both scattering and elevation angle are known, the shape of the gas-liquid interface can be inferred from the positions of the glare points relative to the contour of the droplet.
In the simple case of an ellipsoidal cap shaped droplet the aspect ratio and rotational angle of the droplet can be inferred analytically from the system of the three glare point positions relative to the contour of the droplet \cite{Dreisbach2023}.
However, for the highly deformed gas-liquid interface of a droplet during impingement to position of the glare point is highly non-linear, as the glare points split into two and possibly multiple glare points, which can further split or coalesce depending on the curvature of the gas-liquid interface, as can be seen in Figure~\ref{fig:comp_render_real}.
Consequently, the existence of an analytical solution is unlikely, which further warrants the employment of neural network techniques for the volumetric reconstruction.

\subsection{Synthetic data acquisition}
\label{subsec:synthetic_data}

In order to circumvent the aforementioned matching problem between simulation and experiment, the image data used for training the neural network was acquired by means of synthetic image generation.
For this purpose the render program \textit{Blender} with the \textit{LuxCore} package for physically correct ray-tracing was used.

It is well known that neural networks trained on synthetic data exhibit a reduced accuracy once employed on real data.
This performance gap is dependent on the difference between the feature distributions of the real and synthetic datasets \cite{Csurka2017,Shrivastava2017}.
Therefore, it is crucial that the synthetic training images resembles the real images from the experiment as closely as possible.
With this scope the optical setup of the experiments was accurately recreated in the virtual rendering environment.
The light sources were modelled as unpolarised LED lights with a single wavelength and placed at the according positions that match the scattering and elevation angles of the experiments.
The divergence angle of the light was set to $ 4^\circ$ in order to match the experiments.
The intensities of the lateral light sources, as well as the backlight were tuned empirically to be in accordance with the images observed in the experiments.
The objective lens of the camera was represented by a plano-convex aspherical lens with $60$\,mm focal length an aperture in the shape of a regular pentagon with a circumdiameter of $D_A=10.6$\,mm.

A \textit{Python}-script was set up to import the droplet geometries into this setup and optionally apply transformations through rotation and scaling before rendering a synthetic image of the droplet.
The required geometry of the gas-liquid interface was extracted from the results of direct numerical simulations (DNS) of droplet impingement conducted by Fink \citet{fink2018} with the phase-field method.
The simulations covered droplet impingement on flat substrates, resulting in asymmetrical droplet deformation and structured surfaces, which resulted in an anisotropic wetting of the substrate and non-asymmetrical droplet deformation.
The surface structure consisted of regular square grooves that have a width, height and spacing of $60$\,$\mu m$ and matched the experiments.
The fluid properties, as well as the kinematic conditions of the simulations matched the experiments, in order to acquire representative training data.

For each time step in the simulations an image was rendered with the blender pipeline and in the case of non-asymmetrical droplet deformation multiple images were rendered at different observation angles.
In order to simulate the droplet impingement on an arbitrarily rotated structured surface the droplet was rotated around the vertical axis in $10^\circ$ increments for a total of $360^\circ$.
The resulting synthetic RGB-shadowgraphy images are shown in Figure~\ref{fig:comp_render_real}(b) in comparison to real images from the experiment for a similar physical time after droplet impingement.

\subsection{Neural network training}
\label{subsec:network_training}

For the volumetric reconstruction of the gas-liquid interface a state-of-the deep learning method based on the concept of the Pixel-aligned Implicit Function (PIFu) \cite{saito2019} is used.
The core concept of PIFu is the implicit representation of the three-dimensional topology as a level-set function through a neural network.
In this approach a multi-layer perceptron (MLP) predicts the three-dimensional occupancy field for various locations in the image plane by probing a defined set of distances in the out-of-plane coordinate for the prediction whether the point coordinate lies within or outside of the predicted geometry.
The input to the MLP are image features extracted by a convolutional neural network (CNN), in particular the so-called hourglass network \cite{Newell2016}, which subsequently down- and upsamples the input image.
The resulting feature maps extracted are therefore pixel-aligned, i.e. each pixel position of the extracted feature maps has an accordingly aligned area in the input image. 
Furthermore, the relation between distant locations in the image can expressed by the pixel-aligned feature maps, thus enabling the neural network for both global and local reasoning.

The coupled neural networks were trained jointly by supervised learning with the RMSProp optimisation algorithm \cite{Tieleman2012}, which is an extension of stochastic gradient descent with momentum (SGDM) \cite{Qian1999}.
The training hyperparameters were originally chosen according to the original publication \cite{saito2019}.
However, it was found that a reduction of the training iterations led to better results for the considered dataset.
Therefore, the networks were trained for eight epochs on $37.300$ training samples, with a batch size of $12$ and an initial learning rate of $0.001$ that is reduced by a factor of ten after epochs four and six.

\subsection{Volumetric reconstruction}
\label{subsec:volumetric_reconstruction}

The neural network learned on droplet dynamics is employed for the volumetric reconstruction on the basis of synthetic, as well as real images from experiments.
The evaluation on synthetic data allows for an accurate evaluation of the reconstruction accuracy, as 3D ground truth data is available, while the reconstruction of experimental images is used to evaluate the generalisability of the network to the real world task.
In order to provide the neural network with the appropriate input data the images recorded in the experiment are first pre-processed by the following steps.
First, the images are scaled and cut into the expected input format of the neural network of $512$\, x $512$\,px.
Afterwards a color correction is performed in order to correct for the effects of polychromatic light and cross talk between the camera channels, according to the method described in \cite{Dreisbach2023}.
In the resulting image each color channel only represents the response of the camera to the respective lights source, i.e the red color channel only shows the illumination of the droplet from the red light source, etc.

\begin{figure}[h]
	\centering
        \includegraphics[width=1.0\columnwidth]{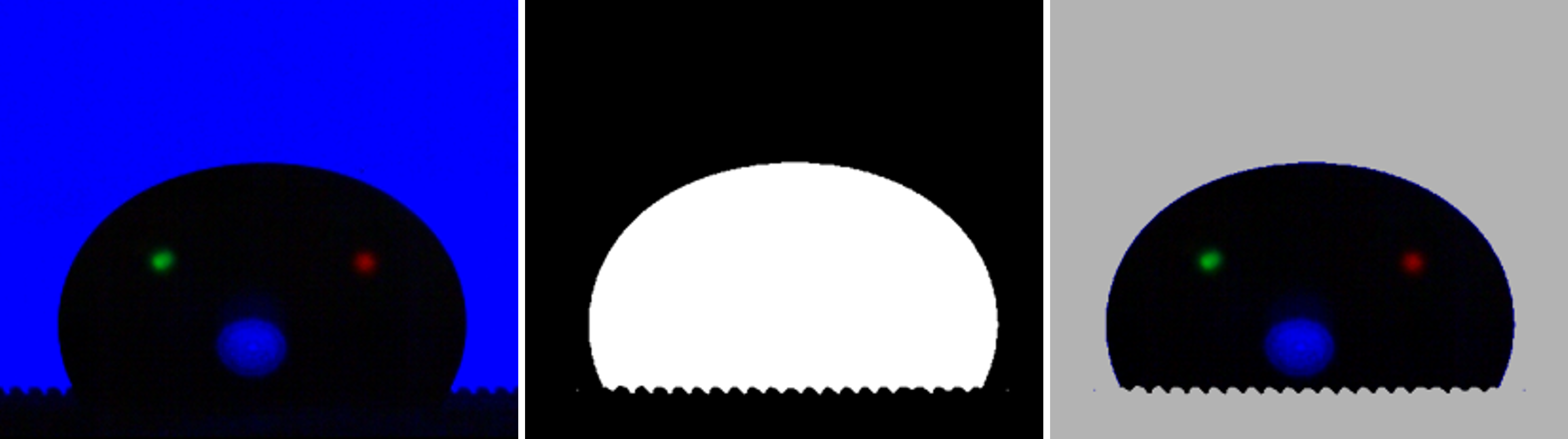}
	\caption{Image of a droplet that is deposited on a structured substrate (left), binary mask (middle) and masked image (right).}
	\label{fig:masking}
\end{figure}

In the final pre-processing step a binary image mask is obtained from the shadowgraph contour of the droplet, which serves as an additional input to the neural network.
In order to allow for the distinction between the contours of the droplet and substrate, first the outline of the substrate is determined from an initial image, in which the droplet is not yet in frame.
Afterwards the substrate contour is removed from all later masks, and thus the information of the solid-liquid interface is imposed on each frame, as evident from Figure~\ref{fig:masking}~(middle).
Subsequently, the input images are overlayed with their respective mask, as can be seen in Figure~\ref{fig:masking}~(right) and inputted to the neural network in order to obtain the frame-wise volumetric reconstruction of the droplet during impingement.

\subsection{Evaluation metrics}
\label{subsec:eval_metrics}

\begin{figure}[h]
	\centering
	\subfigure[Image of the spheroidal droplet before impingement]
        {\includegraphics[width=0.5\columnwidth]{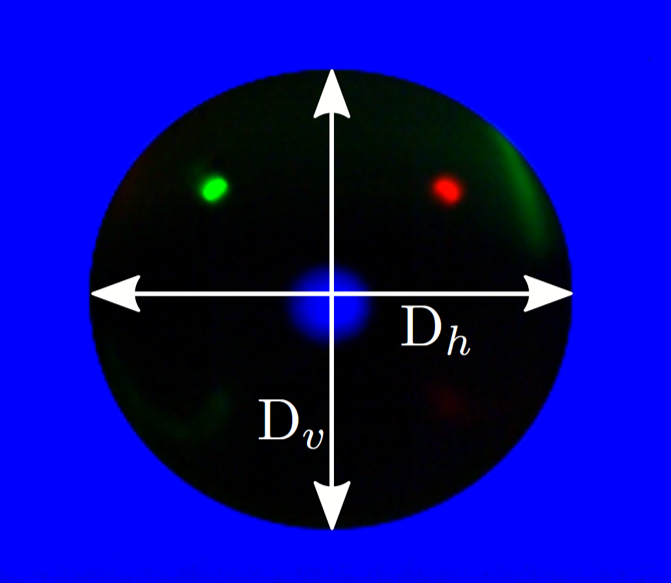}
        \label{fig:spheroid}}
	\hfill
	\subfigure[Intersection (top) and union area (bottom)]
        {\includegraphics[width=0.4\columnwidth]{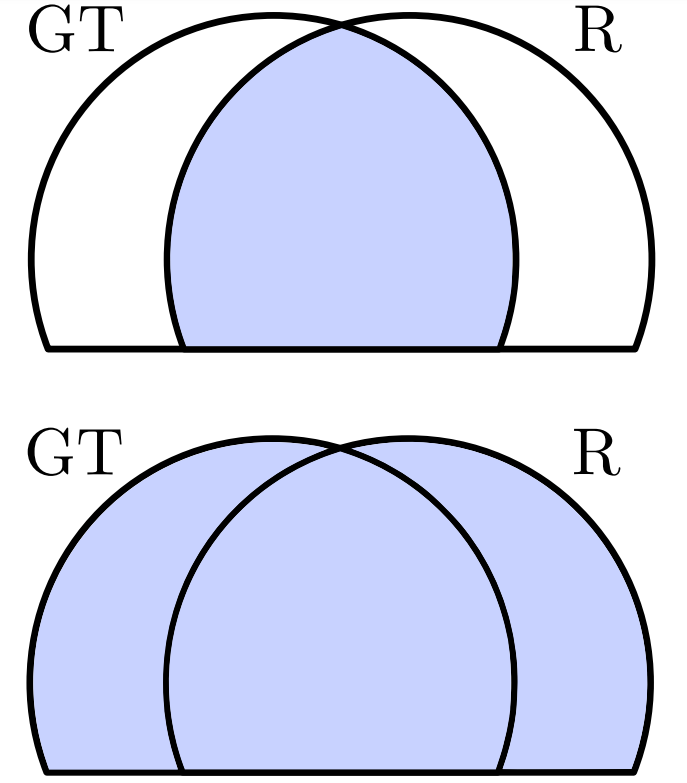}
        \label{fig:3D-IOU}}
	\caption{(a) Image of a spheroidal droplet before impact with markers indicating the horizontal $D_h$ and vertical semi-axis $D_v$ of the ellipse fitted in the image plane; (b) Intersection and union areas of the reconstruction $R$ and ground truth $GT$ in two-dimensional space.}
\end{figure}

The performance of the neural network is evaluated considering the reconstructed three-dimensional shapes and the availability of ground truth data.
The following metrics are used for the evaluation:

\begin{itemize}
    \item The three-dimensional intersection over union $3D-IOU = \frac{R \cap GT}{R \cup GT}$ is calculated as the fraction of the intersection volume between the reconstruction $R$ and ground truth $GT$ and the union volume of $R$ and $GT$.
    The $3D-IOU$ is a straight-forward extension of 2D $IOU$ \cite{Everingham2010} (see Figure~\ref{fig:3D-IOU}) to three dimensions and therefore provides a measure for the spatial accuracy of the reconstruction in 3D-space.

    \item The bias error of the reconstructed volume $\delta_V$ is calculated by the absolute deviation of the mean volume of the reconstructed shapes $\overline{V} = \frac{1}{n}\sum_{i=1}^{n} V_{i}^{'} $ from the ground truth volume $V_{GT}$, and given relative to the ground truth volume \mbox{$\delta_V = |\frac{V_{GT} - \overline{V}}{V_{GT}}|$} \cite{Bendat2010}.
    
    \item The measured uncertainty of the reconstructed volume $\sigma_V$, is calculated by the standard deviation of the errors between the reconstructed volume and the ground truth volume, and given relative to the ground truth volume \mbox{$\sigma_V = \frac{1}{V_{GT}} \sqrt{ \frac{1}{n}\sum_{i=1}^{n} (V_{i}^{'} - \overline{V})^2}$} \cite{Bendat2010}.
\end{itemize}

The reconstruction of the synthetic validation dataset is evaluated by means of the $3D-IOU$.
The results can be interpreted as the baseline performance of the neural network, that can be reached for a perfect agreement of synthetic training data with real experimental data.
The reconstruction of experimental images is evaluated by means of the volumetric bias error $\delta_V$ and uncertainty $\sigma_V$ of the reconstruction relative to the integral ground truth volume $V_{GT}$.
The ground truth volume in the experiments is estimated from the shadowgraph contour of the droplet before its impact on the surface, as opposed to the synthetic test case no 3D ground truth is available.
The shape of the droplet is assumed to be spheroidal with axisymmetry around the vertical axis, and consequently, the volume can be calculated as $V_{GT} = \frac{4}{3} \pi D_s^3$ with the equivalent spherical diameter $D_s=\sqrt[3]{D_h^2 D_v}$.
$D_h$ and $D_v$ are the horizontal and the vertical semi-axis of the falling droplet, as indicated in Figure~\ref{fig:spheroid} that are determined by an ellipse fit to the shadowgraph contour with the method of Taubin \cite{Taubin1991}.
The ground truth volume is averaged over all frames of the undeformed droplet before impact (usually eleven frames), in order to reach a high accuracy of the estimation for the ground truth volume.
The uncertainty of the ground truth volume, measured by the standard deviation over all consecutive frames amounts to $\sigma_{V,GT} = 0.06\%$ of $V_{GT}$ and is therefore negligible.
Furthermore, the reconstruction is qualitatively evaluated by the comparison of the contours of the reconstructed droplet shapes and the shadowgraph images.

\section{Results}
\label{sec:results}

In the following section, first the results of synthetic image generation by means of the rendering are presented.
This is followed by a validation of the neural network through the reconstruction of synthetic images and a quantification of the reconstruction accuracy in subsection~\ref{sec:res_val}.
In subsection~\ref{sec:res_flat} the results for the reconstruction of droplet impingement on flat substrates, characterised by axisymmetrical droplet deformation, are presented.
Finally, the reconstruction of droplet impingement on structured surfaces with different degrees of anisotropy in wetting, resulting in non-axisymmetrical droplet deformation is evaluated in subsection~\ref{sec:res_structured}.

\subsection{Synthetic image rendering}
\label{sec:res_rendering}

\begin{figure}[h]
	\centering
        \includegraphics[width=1.0\columnwidth]{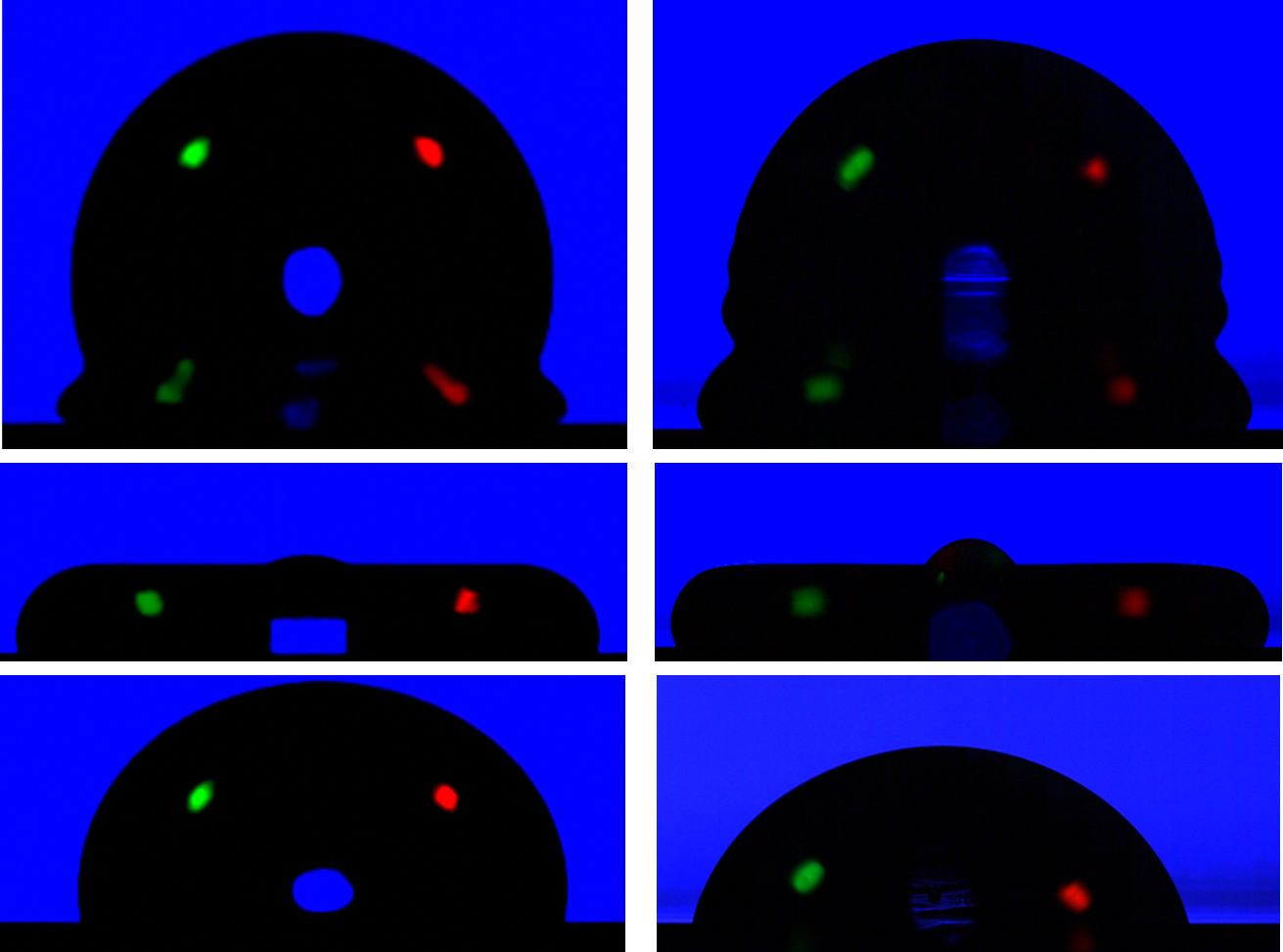}
	\caption{Comparison of synthetic images (left) and recordings from the experiment (right) at similar time steps.}
	\label{fig:comp_render_real}
\end{figure}

Synthetic images were generated by means of the rendering pipeline in \textit{Blender} from the three-dimensional gas-liquid interface for each time step extracted from the results of the direct numerical simulation of droplet impingement conducted by Fink~\citet{fink2018}, as described in subsection~\ref{subsec:synthetic_data}.
The simulation considered the perpendicular impingement of a $d_0=2.1$\,mm water droplet at an impact velocity of $u_0=0.62$\,m/s on a flat hydrophobic Polydimethylsiloxane (PDMS) substrate that resulted in an axisymmetrical droplet deposition.
The resulting synthetic RGB-shadowgraphy images are shown in Figure~\ref{fig:comp_render_real}~(left) in comparison to real images from the experiment for a similar physical time after droplet impingement.
The experiments were conducted with the setup introduced in subsection~\ref{subsec:experimental_setup} and featured the impingement of a $d_0=2.08$\,mm water droplet at an impact velocity of $u_0=0.7$\,m/s on a flat hydrophilic silicon oxide (SiOx) substrate.
It should be noted that while the impact parameters are not identical, they similar enough to allow for phenomenological comparison of the obtained images.

As can be seen in Figure~\ref{fig:comp_render_real} the position and appearance of the glare points, as well as the focus of the shadowgraph contour and the glare points was reproduced accurately in the the synthetic images.
The illumination of the lateral green and red light sources resulted only in $p=0$ glare points in the rendering, which agrees well with the experiments and theory \cite{Dreisbach2023}.
Furthermore, the blue $p=1$ glare point from background illumination is reproduced in the rendering.
The good agreement between synthetic and experimental images demonstrate the capability of the rendering approach for realistic synthetic data generation and thus validates the assumptions made in the setup of the render environment.

Note that the complex deformed shape of the gas-liquid interface right after impact produces multiple glare points, as each capillary wave with an appropriate surface angle casts a glare point.
These glare points split and merge over time due to the temporal development of the interface deformation, which leads to a highly non-linear behaviour of the color-coded glare points.
Consequently, the employment of deep learning techniques for the reconstruction of deformed gas-liquid interface from these glare points is suggested.

\subsection{Validation on synthetic image data}
\label{sec:res_val}
The PIFu neural network \cite{saito2019} was learned for the volumetric reconstruction of the gas-liquid interface during droplet impingement by training on labelled datasets of synthetically rendered images, as described in section~\ref{sec:methodology}.
The input images were rendered from the gas-liquid interface extracted from $1.479$ time steps of numerical simulation by Fink~\citet{fink2018}.
Of these, $1.014$ snapshots originated from the simulation of droplet impingement on a structured PDMS surface that resulted in a complete rebound and non-axisymmetrical droplet deformation, while the remaining $465$ snapshots feature droplet impingement on a flat substrate resulting in axisymmetrical droplet deposition.
In the following the trained state of the PIFu neural network for the reconstruction from RGB-shadowgraphy images is referred to as \textit{DFS2023C}.
Additionally, a second synthetic image dataset without glare points, and therefore only featuring the shadowgraph contour, was generated from the same ground truth data.
A second benchmark version of the neural network was trained on the dataset without glare point and is referred to as \textit{DFS2023E} in the following.
Both datasets were split in a ratio of $70/10/20$ into a training dataset and separate validation and testing datasets that are not used directly for the training of the network and can therefore be used to evaluate the reconstruction accuracy on unknown data. 
The split of the dataset was performed at the level of time steps, so that all $36$ images rendered at different observation angles for each respective time step are either completely used for training or validation, in order to conserve a clear distinction between training and validation data.
During training the reconstruction performance was evaluated by the three-dimensional Intersection over Union ($3D-IOU$, see~\ref{subsec:eval_metrics}) on the training and validation datasets.

\begin{figure}[ht]
    \centering
    \includegraphics[width=1.0\columnwidth]{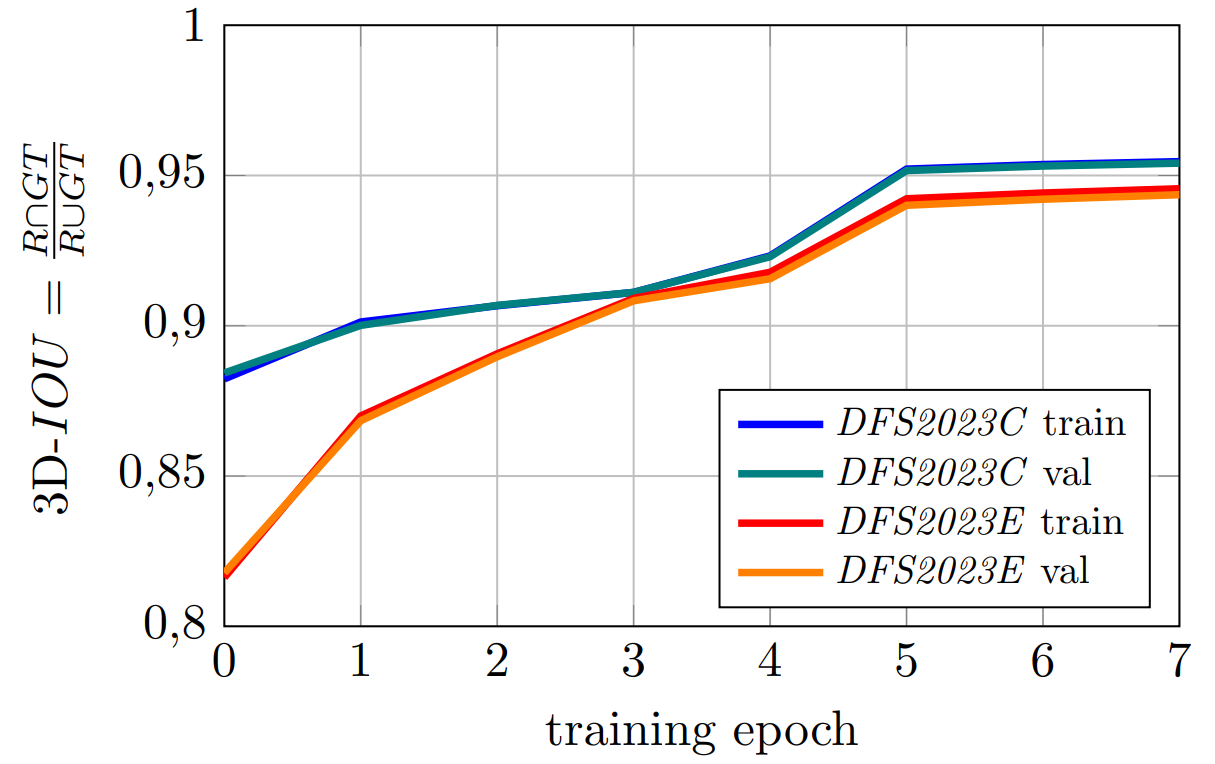}
    \caption{3D-$IOU$ during training of the \textit{DFS2023C} and \textit{DFS2023E} networks on the training and validation datasets.}
    \label{fig:3D-IOU_train_val}
\end{figure}

As can be seen in Figure~\ref{fig:3D-IOU_train_val} the reconstruction accuracy improves over time on both the training and validation data set, which indicates a successful learning of the neural network.
In particular the strictly monotonically increasing validation performance indicates that the network did not overfit the training data, but rather that a generalisation capability to unknown data was maintained.
However, the training and validation performance is very similar, which likely results from both data stemming from the same distribution with a fine time temporal resolution, so that a close interpolation of the unknown time steps by the neural network is possible.
Precisely, \textit{DFS2023C} reaches a performance of $3D-IOU=0.954$ on the validation dataset at the end of the training and $3D-IOU=0.955$ on the training dataset, while \textit{DFS2023E} reaches $3D-IOU=0.944$ and $3D-IOU=0.946$, respectively.
Therefore, the reconstruction accuracy appears to be elevated with glare points.
In particular, considering the error relative to the perfect reconstruction result ($3D-IOU=1$) the difference is significant with a $18.6\%$ lower relative error for \textit{DFS2023C}.
Consequently, the results on synthetic data suggest that glare points facilitate the learning of the three-dimensional droplet shape.

\subsection{Reconstruction of symmetrical experimental data}
\label{sec:res_flat}

In the following, the results of the volumetric reconstruction on the basis of images obtained experimentally by means of RGB-shadowgraphy is evaluated.
The experiments were conducted with the test rig described in subsection~\ref{subsec:experimental_setup} and the recorded images were pre-processed according to subsection~\ref{subsec:volumetric_reconstruction}.
Furthermore, prior to the reconstruction all images were cropped uniformly in order to reach a high resolution in the reconstruction, while conserving the scale between time steps.
The experiments covered the impingement of water droplets with an equivalent diameter of $d_0=2.08$\,mm onto a flat hydrophilic SiOx surface with an impact velocity of $u_0=0.7$\,m/s.
The outcome of the impact was a deposition of the droplet with axisymmetrical deformation.
Both states of the neural network trained on simulation data with anisotropic wetting, \textit{DFS2023C} and \textit{DFS2023E}, were employed.
Additionally a benchmark model \textit{DF2022} was learned solely on the part of the training dataset with glare points that comprised of isotropic wetting of the flat substrate and therefore had strictly axisymmetrical deformation.

\begin{figure}[htbp!]
\centering
    \includegraphics[width=1.0\columnwidth]{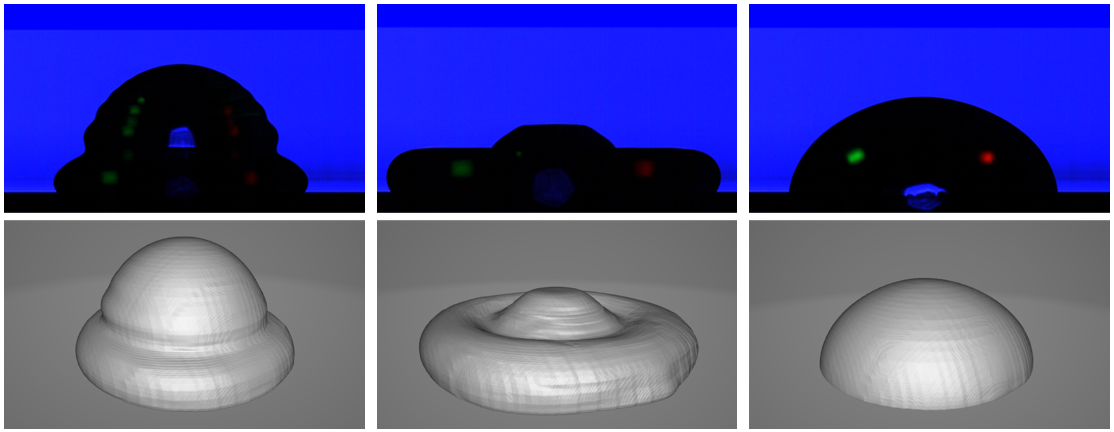}
    \caption{Reconstructed three-dimensional gas-liquid interfaces (bottom) and according input images (top) for droplet impingement on the flat hydrophilic SiOx-substrate.}
    \label{fig:rec_flat}
\end{figure}

\begin{figure}[ht]
    \centering
    \includegraphics[width=1.0\columnwidth]{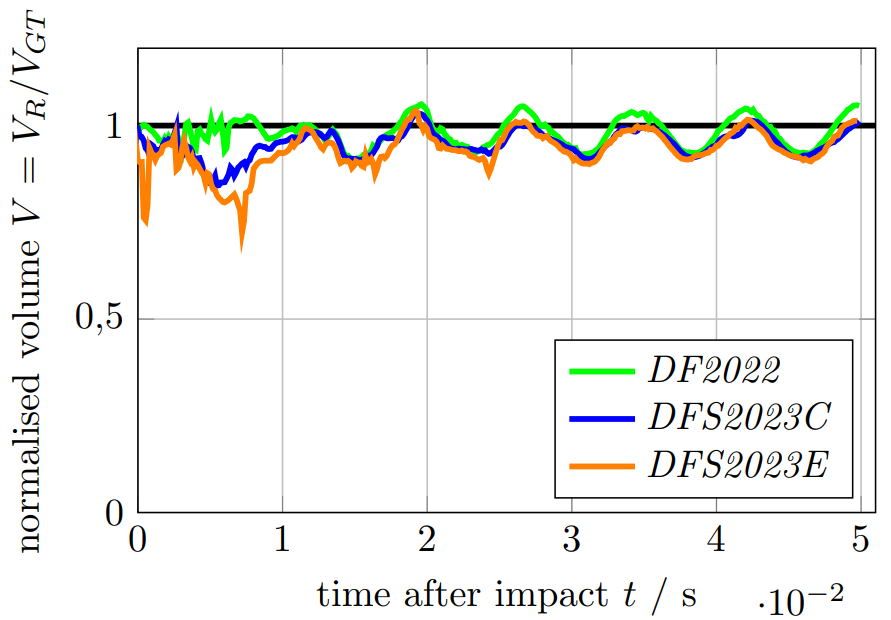}
    \caption{Temporal development of the normalised integral volume of the reconstruction for droplet impingement on the flat SiOx substrate with \textit{DF2022}, \textit{DFS2023C} and \textit{DFS2023E}.}
    \label{fig:vol_flat}
\end{figure}

\begin{figure}[htbp!]
\centering
    \includegraphics[width=1.0\columnwidth]{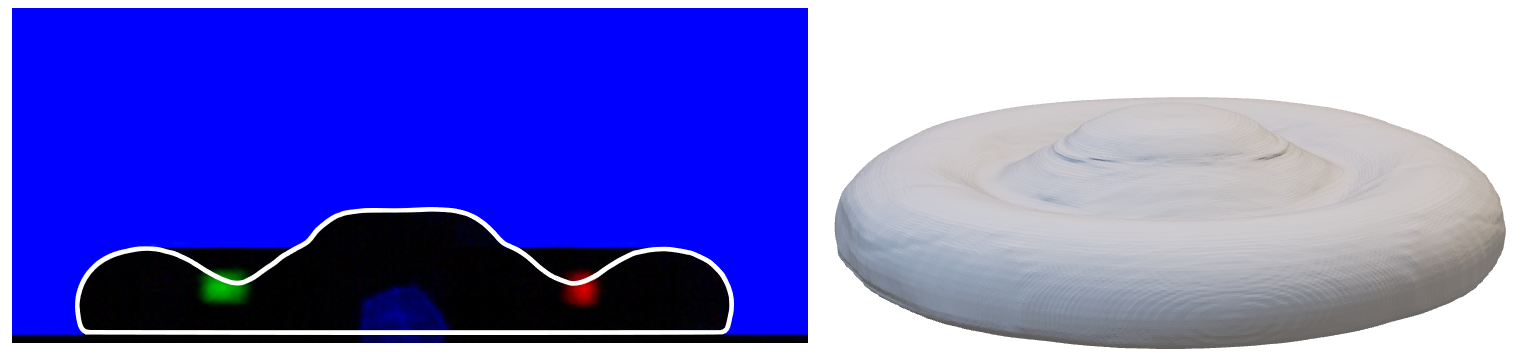}
    \caption{Reconstruction of a frame with self-occlusion of the gas-liquid interface (right) and respective input image (left) overlaid with the contour of the reconstruction.}
    \label{fig:rec_occlusion}
\end{figure}

In Figure~\ref{fig:rec_flat} three exemplary three-dimensional gas-liquid interface geometries reconstructed by the \textit{DF2022} network are illustrated with their respective input images.
As can be seen, the network was able to infer physically reasonable shapes from the input images that align well with the contour of the input image.
Furthermore, the reconstruction results are axisymmetric on the global scale, thus revealing that the rotational symmetry that is characteristic for a droplet impingement on flat substrates was learned well by network.
The results of the \textit{DFS2023C} and \textit{DF2022} networks deviate slightly more from axisymmetry than those of \textit{DF2022}, which indicates an influence of the training data on the learning of symmetries.

Figure~\ref{fig:vol_flat} shows the temporal development of the integral volume $V_R$ of the reconstruction relative to the ground truth volume measured from the images in the experiment $V_{GT}$, as described in section~\ref{subsec:eval_metrics}.
It can be observed that all versions of the PIFu network were able to reconstruct the volume of the gas-liquid interface with good agreement to the experiment.
In particular, the \textit{DF2022} network reached a close agreement indicated by a low bias error of $\delta_V = 1.8\%$.
In comparison, the version trained on simulations including non-axisymmetrical wetting produced higher bias errors in the reconstruction, with $\delta_V = 4.7\%$ for \textit{DFS2023C} and $\delta_V = 6.2\%$ for \textit{DFS2023E}.
The uncertainty of the reconstructed volume is similar across all versions of the network, with $\sigma_V = 3.7$ for \textit{DF2022}, $\sigma_V = 3.5$ for \textit{DFS2023C} and $\sigma_V = 5.1$ for \textit{DFS2023E}.
Overall the networks trained on synthetic image data with glare points reached a lower bias error and uncertainty in the reconstruction in comparison to the model trained without glare points, even though the droplet dynamics were axisymmetrical and thus sufficiently represented by the shadowgraph contour.
Furthermore, it was found that the error of the reconstructed volume is closely related to the oscillation of the droplet after impact, which can directly be observed by the oscillation of the reconstructed volume.

In Figure~\ref{fig:rec_occlusion} the reconstruction result for an image frame with self-occlusion of the gas-liquid interface is shown.
The droplet contour in the shadowgraph image appears to have a sharp corner between the higher central region and the flat lamella surrounding it, however in reality the gas-liquid interface is smooth and the most outer part of the lamella is raised higher than the section closer to the central peak.
The white line overlaid with the input image on the left indicates the the gas-liquid interface of the 3D-reconstruction for the cross section in the image plane.
As can be seen, the smooth contour of the gas-liquid interface, as well as the large curvature that should occur are successfully estimated by the neural network in a physically correct manner for the regions that are obscured in the input image.
The reconstruction of occluded regions within the three-dimensional shape reveals the strength of the deep learning approach to interpolate large unknown regions respecting the underlying physics contained in the numerical training data.

\subsection{Reconstruction of non-asymmetrical experimental data}
\label{sec:res_structured}

\begin{figure}[ht]
    \centering
    \includegraphics[width=0.85\columnwidth]{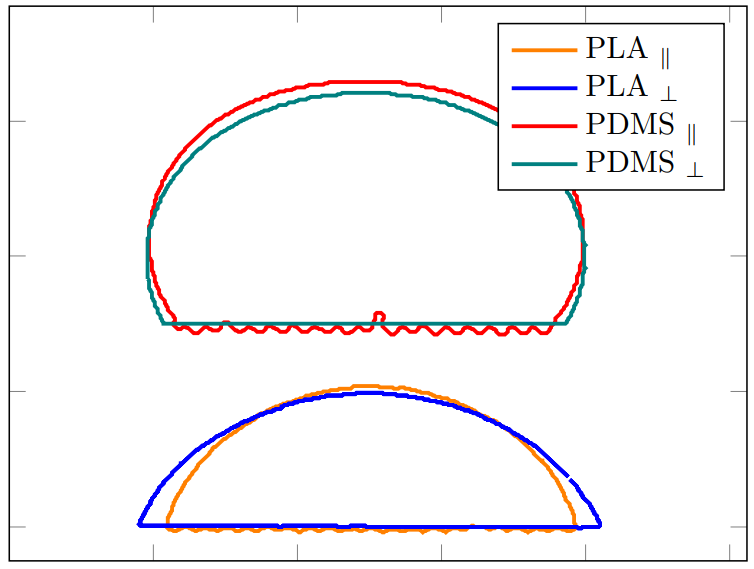}
    \caption{\textit{Contours of the shadowgraph images for deposited droplets on the structured PLA and PDMS substrates at parallel and transversal viewing orientation}}
    \label{fig:exp_contour}
\end{figure}

In the following section the volumetric reconstruction on the basis of experiments of water droplet impingement on two different structured substrates, in particular hydrophilic polylactide (PLA) and hydrophobic Polydimethylsiloxane (PDMS), is presented.
Images showing the structure of the both substrates can be found in Figure~\ref{fig:substrates} in the Appendix.
The anisotropic wetting of structured surfaces leads to non-axisymmetrical droplet deformation during impingement, with a larger spreading ratio and lower contact angles in the direction parallel to the grooves in comparison to the transversal direction.
The resulting static shape of the droplet after impact on the PLA and PDMS substrates in parallel and transversal direction is illustrated in Figure~\ref{fig:exp_contour}.
Due to the non-axisymmetrical deformation of the gas-liquid interface, the information from only a shadowgraph projection is not sufficient for a volumetric reconstruction.
Consequently, the additional three-dimensional information of the gas-liquid interface encoded in the glare points has to be exploited by the neural network to achieve an accurate reconstruction.
The results of the reconstruction for \textit{DFS2023C} and \textit{DFS2023E} are compared in order to evaluate effectiveness of glare points for the reconstruction of asymmetric droplet deformation.  
Different observation angles, i.e. the angle between the orientation of structures and the camera axis, are considered and the results of the volumetric reconstruction are compared in order to determine the influence of the observation angle on the reconstruction accuracy. 
In the following for both substrates an observation angle of $\alpha=0^\circ$ denotes parallel alignment of camera and substrate, while $\alpha=90^\circ$ denotes a perpendicular alignment.

In the first series of experiments the impingement dropleta on a hydrophilic 3D-printed substrate, that was produced from polylactide (PLA) by Fused Deposition Modeling (FDM), is reconstructed volumetrically.
A black filament was chosen in order to minimise the reflection of the lateral illumination on the substrate that would result in unwanted additional glare points that could disturb the reconstruction.
The pattern of the stacked layers resulting from the 3D-printing process is horizontally aligned with the substrate surface, forming a wave pattern with a characteristic length of $154 \mu$m.
The experiments featured the impingement of water droplets with an equivalent diameter of $D_0=2.27$\,mm at an impact velocity of $u_0=0.45$\,m/s, that was recorded at $0^\circ, 45^\circ, 90^\circ$ orientation angles.
In the experiments an equilibrium contact angle of $\theta_a = 76^\circ$ in the parallel direction and $\theta_a = 63^\circ$ in the transversal direction was measured.
The dynamic contact angles for both directions are found in Table~\ref{tab:dyn_contact_angles} in the Appendix.

\begin{figure}[ht]
    \centering
    \includegraphics[width=1.0\columnwidth]{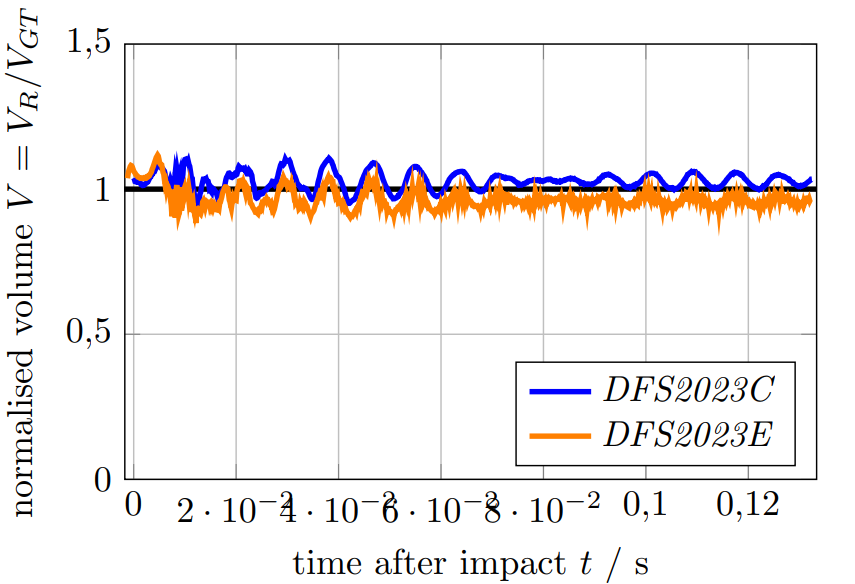}
    \caption{Temporal development of the normalised integral volume of the reconstruction for droplet impingement on the structured PLA-substrate at $\alpha=45^\circ$ observation angle with \textit{DFS2023C} and \textit{DFS2023E}.}
    \label{fig:vol_fdm_45}
\end{figure}

The results of the volumetric reconstruction from experimental images of droplet impingement on the PLA-substrate indicate that both states of the network trained with and without glare points successfully reconstructed the non-axisymmetrical droplet deformation.
\textit{DFS2023C} reached an uncertainty of $5.7\%$ and a bias error of $4.9\%$ averaged over all evaluated observation angles, compared to an uncertainty of $6.0\%$ and bias error of $8.0\%$ of the reconstruction by \textit{DFS2023E}.
Consequently, the training with glare points resulted in lower errors for the reconstruction of non-axisymmetrical droplet dynamics.
More detailed results can be found in Tables~\ref{tab:uncertainties_all} and \ref{tab:bias_all} in the Appendix.

The temporal development of the integral volume for the reconstructed droplet from the $45^\circ$ observation angle is illustrated in Figure~\ref{fig:vol_fdm_45}.
As can be seen \textit{DFS2023C} overestimates the volume of the droplet, while \textit{DFS2023E} does underestimate it.
Furthermore, the oscillation of the droplet is visible in the error of the reconstruction volume as a low frequency oscillation, as was already observed for the reconstruction of asymmetric droplet deformation in section~\ref{sec:res_flat}.
Additionally, a high frequency fluctuation can be found in the reconstructed volume for the neural network trained without glare points (\textit{DFS2023E}), which indicates further random errors in the reconstruction. 

\begin{figure}[ht]
    \centering
    \includegraphics[width=0.85\columnwidth]{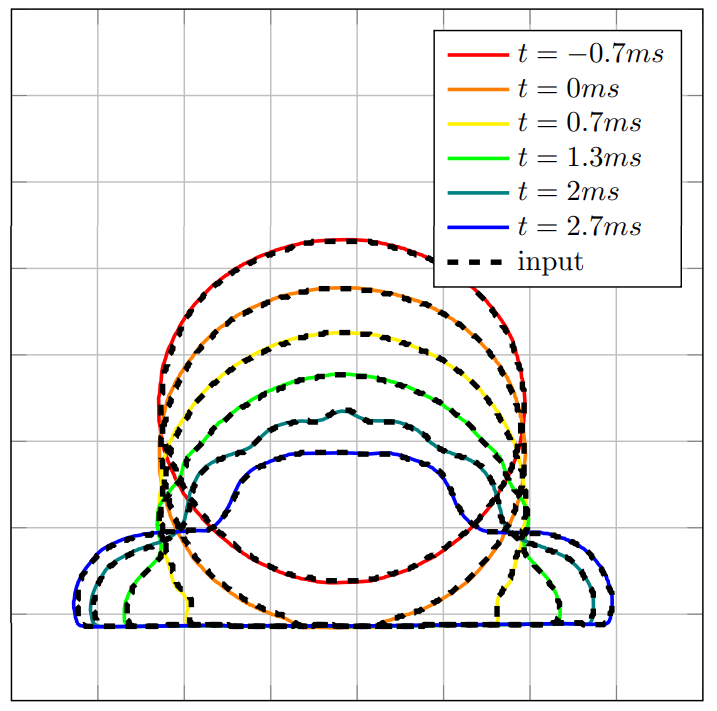}
    \caption{In-plane contour of the reconstructed droplet shapes over time (colored) in comparison to contour of the input shadowgraph (dashed black) for the reconstruction of droplet impingement on the structured PLA-substrate at $\alpha=90^\circ$ by \textit{DFS2023C}.}
    \label{fig:DFS2023C_PLA_90_InP}
\end{figure}

\begin{figure}[ht]
    \centering
    \includegraphics[width=0.85\columnwidth]{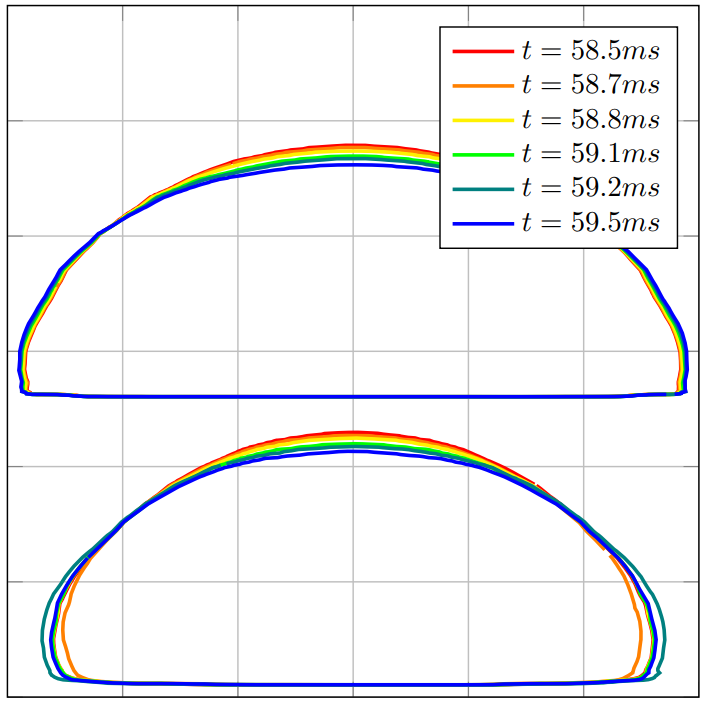}
    \caption{Out-of-plane contour of the reconstructed droplet shapes for consecutive frames for the reconstruction of droplet impingement on the structured PLA-substrate at $\alpha=45^\circ$ by \textit{DFS2023C} (top) and \textit{DFS2023E} (bottom).}
    \label{fig:Contour_OOP_Rec}
\end{figure}

In order to determine the cause for the high frequency oscillation the contour of the 3D-reconstruction is projected in two orthogonal views, in particular the viewing angle matching the input image (in-plane) and a view rotated by $90^\circ$ around the height axis, representing the reconstruction in the depth coordinate (out-of-plane).
Figure~\ref{fig:DFS2023C_PLA_90_InP} shows the temporal evolution of the in-plane contour for the droplet impingement on the PLA-substrate at an observation angle of $45^\circ$ reconstructed by \textit{DFS2023C} in comparison to the contour of the respective input shadowgraph.
As can be seen the in-plane reconstruction reaches a perfect agreement with the input data.
This observation holds true for all other frames of the input sequence and for the \textit{DFS2023E} model as well with very little exception, as can be seen in Figure~\ref{fig:DFS2023E_PLA_90_InP}.
Consequently, the influence of the in-plane reconstruction on the error in the volume can be ruled out.

Figure~\ref{fig:Contour_OOP_Rec} shows the temporal evolution of the out-of-plane contour from the volumetric reconstruction at an observation angle of $45^\circ$ around a time step with particular high volumetric error for both \textit{DFS2023C} and \textit{DFS2023E} (see Figure~\ref{fig:vol_fdm_45}).
The oscillation of the droplet is already significantly dampened around this time instance, as indicated by the minimal movement of the apex of the droplet.
As can be seen, the resulting deformation of the droplet in the out-of-plane coordinate with \textit{DFS2023C} conserves the volume of the droplet.
However, \textit{DFS2023E} reconstructs two of the consecutive time frames with a significantly deviating extent in the out-of-plane coordinate.
For the time step at $t=58.7$ms (orange) the volume of the droplet is underestimated and for $t=59.2$ms (teal) the depth is overestimated, thus explaining the high frequency volume oscillation.
Furthermore, it becomes apparent from the comparison of the reconstruction results of both networks that \textit{DFS2023E} underestimates the out-of-plane extent for all time steps, which is a likely cause for the bias error in the integral volume of the reconstruction.
Consequently, the comparison of the out-of-plane contours for the \textit{DFS2023C} and \textit{DFS2023E} network reveals that the color-coded glare points assist the neural network with the depth estimation during reconstruction.

\subsection{Reconstruction of highly deformed gas-liquid interfaces}
\label{sec:res_pdms}

\begin{figure}[htbp!]
\centering
    \includegraphics[width=1.0\columnwidth]{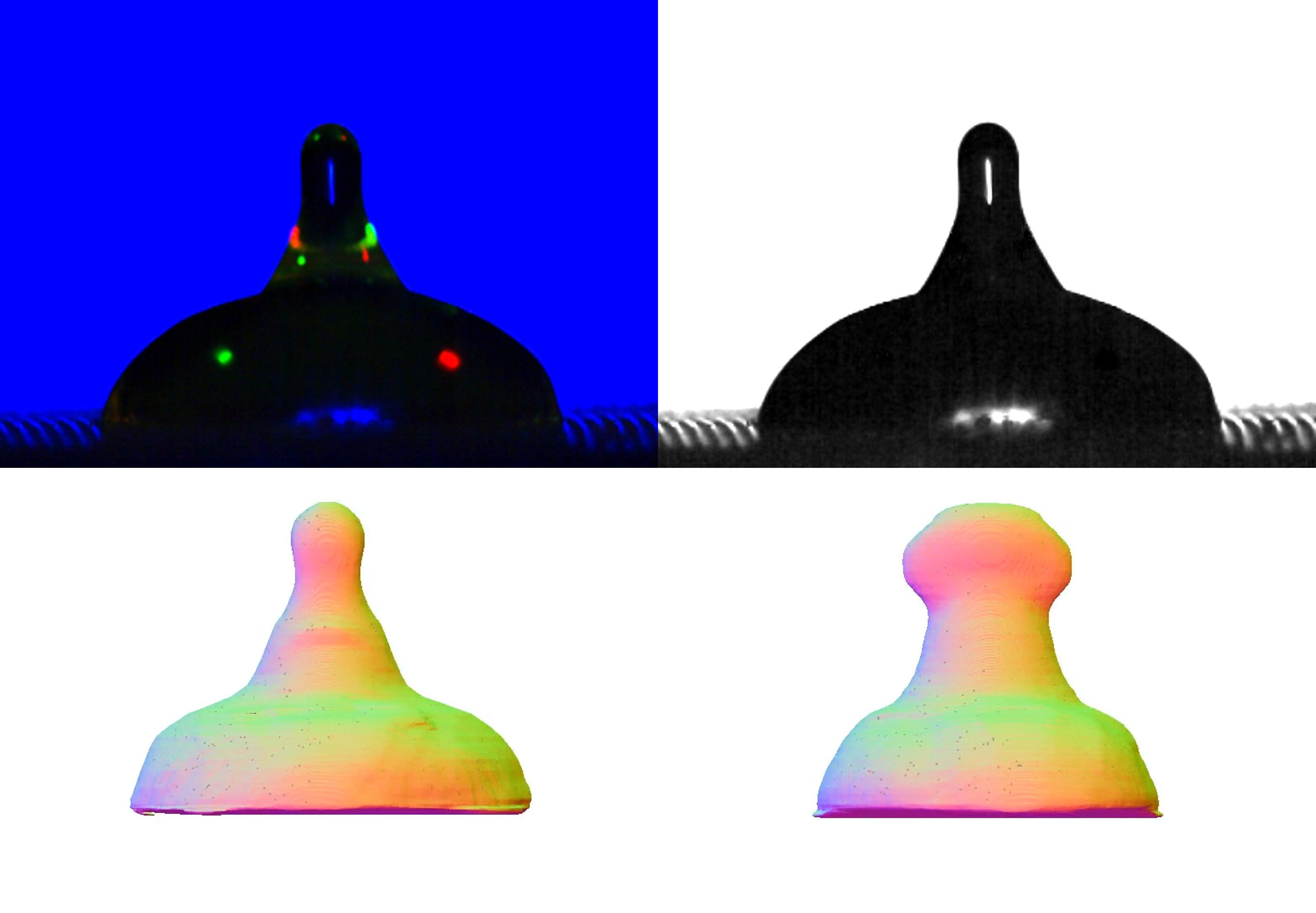}
    \caption{Comparison of the depth estimation for a time step with a high volumetric error during the reconstruction of droplet impingement on the structured PDMS-substrate by \textit{DFS2023C} (left) and \textit{DFS2023E} (right).}
    \label{fig:PDMS_depth_estim}
\end{figure}

In the second series of experiments droplet impingement on the structured PDMS-substrate with regular square grooves that have a width, height and spacing of $60$\,$\mu m$ was reconstructed volumetrically.
The experiments featured the impingement of water droplets with an  equivalent diameter of $D_0=2.26$\,mm on the PDMS-substrate at an impact velocity $u_0=0.88$\,m/s that was recorded at $0^\circ, 45^\circ, 90^\circ$ orientation angles
Since PDMS is a hydrophobic material, its hydrophobicity is further increased by the surface structure, due to the increased surface area \cite{Wenzel1936}.
The equilibrium contact angle was measured to be $\theta_a = 107^\circ$ in parallel and $\theta_a = 97^\circ$ in transversal direction.
The dynamic contact angles in both direction can be found in Table~\ref{tab:dyn_contact_angles} in the Appendix.
The pronounced hydrophobicity of the structured PDMS sample lead to a partial rebound of the droplet in some of the experiments.
The resulting complex three-dimensional deformation of the droplet (see e.g. Figures~\ref{fig:PDMS_depth_estim} and \ref{fig:PDMS_hard_case}) renders a volumetric reconstruction of the gas-liquid interface more difficult.

\begin{figure}[ht]
    \centering
    \includegraphics[width=1.0\columnwidth]{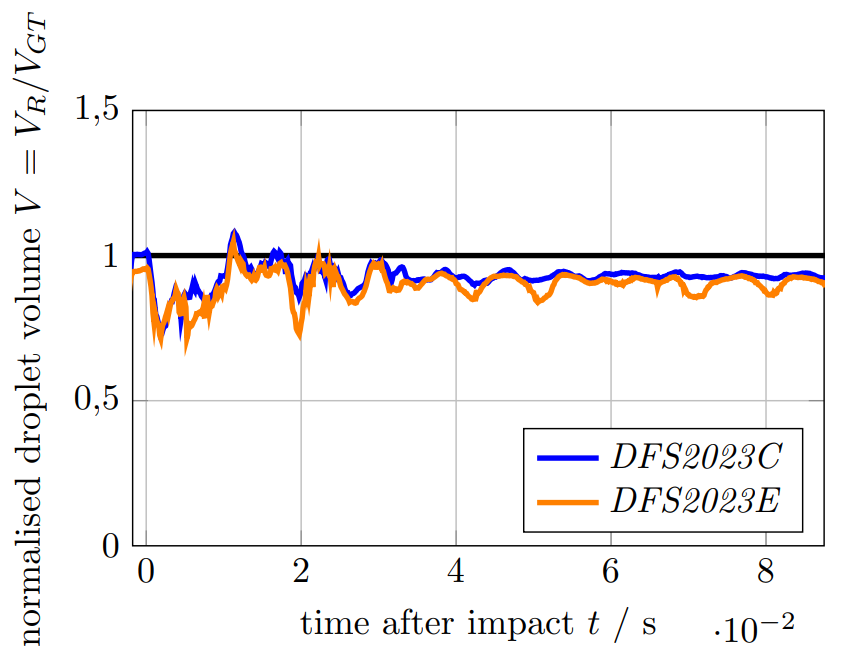}
    \caption{Temporal development of the normalised integral volume of the reconstruction for droplet impingement on the structured PDMS-substrate at $\alpha=90^\circ$ observation angle with \textit{DFS2023C} and \textit{DFS2023E}.}
    \label{fig:vol_pdms_90}
\end{figure}

Figure~\ref{fig:vol_pdms_90} shows the temporal development of the integral volume of the droplet during impingement on the PDMS-substrate reconstructed from images taken at $90^\circ$ orientation angle.
As can be seen both \textit{DFS2023C} and \textit{DFS2023E} successfully reconstruct the three-dimensional dynamics of the significantly deformed gas-liquid interface well.
However, both networks underestimate the volume and the fluctuations during the early stages of the impact with rapid droplet deformation, have a larger magnitude compared to the reconstruction of the lesser deformed droplets during impingement on the PLA-substrate, in particular for the reconstruction by \textit{DFS2023E}.
During the later stages of the droplet impact \textit{DFS2023C} correctly estimates a constant volume, while \textit{DFS2023E} exhibits low frequency oscillation in the reconstructed volume.

\textit{DFS2023C} reached an uncertainty of $6.6\%$ and a bias error of $6.2\%$ averaged over all tested orientations, while \textit{DFS2023E} reached uncertainty of $6.5\%$ and bias error of $8.2\%$.
For both versions of the network the errors are higher in comparison to reconstruction results for the PLA-substrate.
Furthermore, it was found that the reconstruction accuracy is dependent on the observation angle.
The $45^\circ$ observation angle resulted in a significantly lower uncertainty, as well as a lower bias error for the reconstruction of both droplet impingement on PDMS and PLA.
\textit{DFS2023C} reaches a combined uncertainty of $\sigma_V=3.4\%$ and $4.1\%$ bias error, which is significantly lower than the errors in the $0^\circ$ orientation with $\sigma_V=8.5\%$ and $\delta_V=6.7\%$ and $90^\circ$ orientation with $\sigma_V=5.5\%$ and $\delta_V=5.9\%$.
A similar behaviour was observed for the reconstruction with \textit{DFS2023E}.
A detailed summary of the uncertainties and bias errors can be found in Table~\ref{tab:uncertainties_all} and Table~\ref{tab:bias_all}, respectively.

Figure~\ref{fig:PDMS_depth_estim} shows the reconstruction results of both \textit{DFS2023C} and \textit{DFS2023E} for the same image frame recorded in the experiments, however in the input format for both respective networks.
The reconstruction is rotated by $90^\circ$ with respect to the image and therefore shows the depth estimation of the two neural network states.
As can be seen, the complicated shape of the gas-liquid interface was inferred in a physically reasonable way by \textit{DFS2023C}, while \textit{DFS2023E} estimates an unreasonable shape.
This error in the depth estimation can be related to a locally high error in the reconstructed volume, which indicates that the reconstruction of the particular frame proved difficult for the neural networks.
Overall it was found that the depth estimation of \textit{DFS2023E}, in particular for highly deformed droplet shapes, was significantly less consistent in comparison to \textit{DFS2023C}.

\begin{figure}[htbp!]
\centering
    \includegraphics[width=1.0\columnwidth]{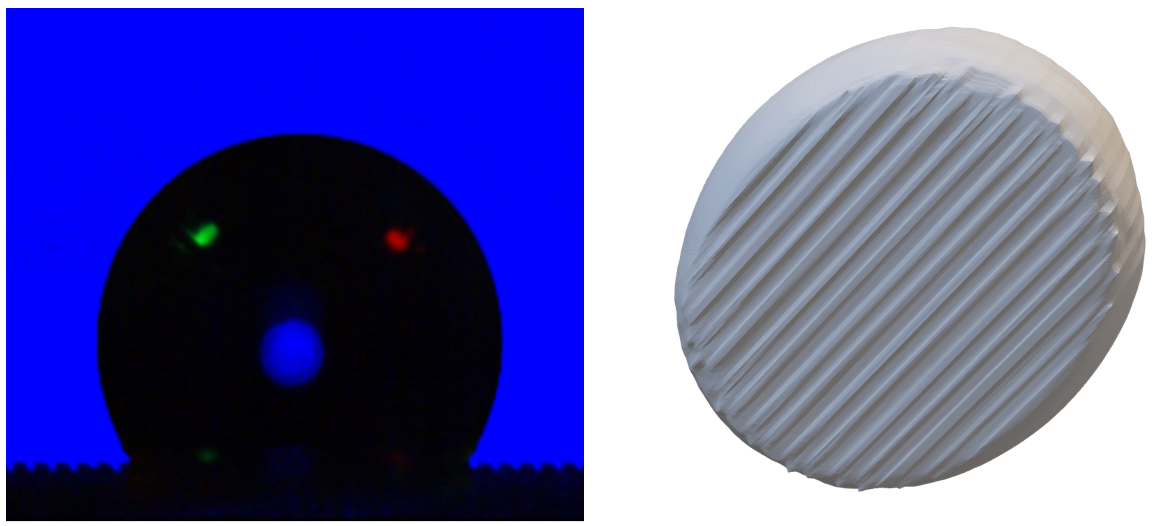}
    \caption{Reconstruction of the droplet during the wetting of the structured surface by \textit{DFS2023C} (right) and the respective input RGB-shadowgraph recorded at $\alpha=0^\circ$ observation angle. Note that the reconstruction is rotated.}
    \label{fig:PDMS_wetting}
\end{figure}

Figure~\ref{fig:PDMS_wetting} shows the reconstruction of an image frame recorded during droplet impingement on the PDMS-substrate at an observation angle of $0^\circ$.
The reconstructed droplet shape is rotated, so that the contact area between the liquid and the solid substrate is visible.
As can be seen, the grooves that are also visible in the input image are extended over the whole depth of the reconstructed volume, thus revealing that small scale features are accurately reconstructed by the neural network.
However, the reconstruction of the wetted state was only possible if the surface structure was visible in the input images and consequently only for the $\alpha=0^\circ$ observation angle.
Under other observation angles the surface appears to be flat in the images and the liquid-solid interface is reconstructed as a flat plane by the neural network.

It should be noted that the training dataset contained both flat and structured surfaces, which could be a source of confusion for the neural network that leads to the reconstruction of flat surfaces if the grooves are not visible. 
In order to determine the effect of training data on the reconstruction of structured surface wetting, a benchmark version of the network \textit{DS2022} is trained solely on the part of the training data that considers the droplet impingement on structured surfaces.
It was found that the \textit{DS2022} network also only reconstructs the wetted state accurately if the grooves of the substrate are visible in the input image, thus indicating that the effect is independent from the training data.


\begin{figure}[htbp!]
\centering
    \includegraphics[width=1.0\columnwidth]{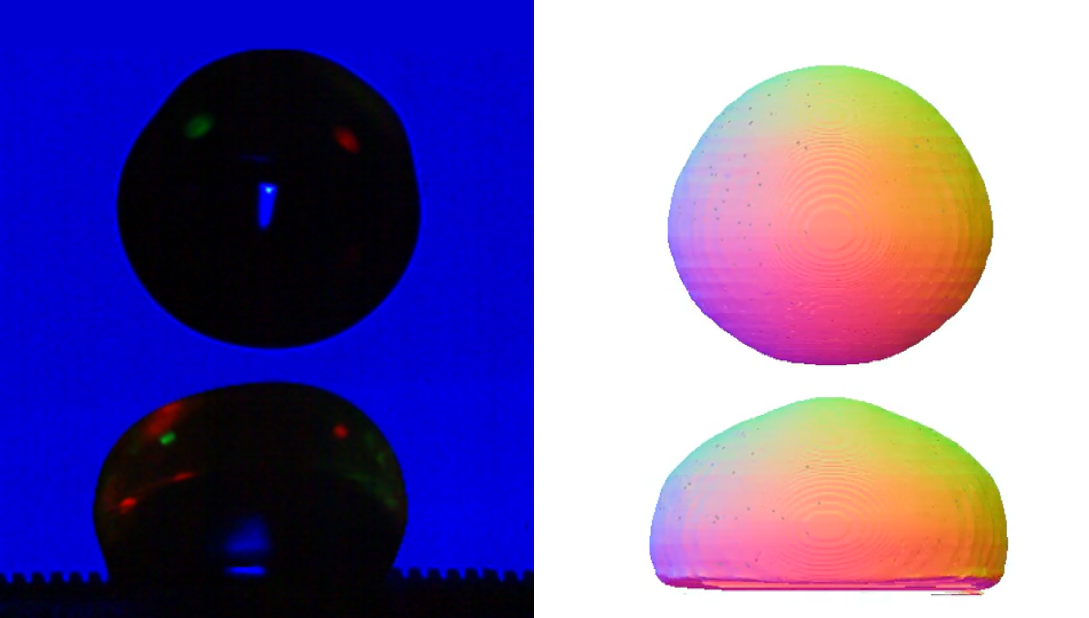}
    \caption{Reconstruction of multiple droplets (right) by \textit{DFS2023C} and the respective input images (left). Note that the reconstruction is rotated by $90^\circ$ relative to the input.}
    \label{fig:topological}
\end{figure}

Figure~\ref{fig:topological} shows a recording from experiment and the respective volumetric reconstruction obtained by the \textit{DFS2023C} network for a drop impact experiment that resulted in a partial rebound.
As can be seen, the volume of multiple droplets was reconstructed successfully.
Furthermore, the separation into two droplets, as well as the subsequent coalescence were both reconstructed accurately by the neural network.
These results demonstrate that topological changes can be accurately represented by the learned neural network.
Other experiments show the emergence of a Worthington jet that leads to the detachment of a tiny droplet at high velocities, which was accurately reconstructed by the network as well.
Both of these droplet impingement phenomena were not represented by the training data and in particular no topological changes were part of the training data, as the droplet always stayed intact.
These results suggest a high flexibility of the neural network approach to the reconstruction of unknown outcomes during droplet impingement.

\section{Discussion}
\label{sec:discussion}

The presented results for the volumetric reconstruction of the droplet dynamics during impingement indicate that the proposed method based on neural networks is able to accurately reconstruct the three-dimensional shape of the gas-liquid interface from a single image.
This is demonstrated by the successful reconstruction of both axis-symmetrical deformation and asymmetric deformation of the droplet due to the impact on flat and structured surfaces, respectively.
In particular, droplet impingement on structured surfaces lead to complex shapes of the gas-liquid interface that pose a considerably more challenging task and therefore reveal the capacity of the data-driven approach for volumetric reconstruction.
The validation on synthetic data shows a good agreement of the reconstruction with the 3D ground truth and thereby reveals that the non-axisymmetrical shape of the droplet resulting from anisotropic wetting of the structured surfaces can be successfully reconstructed by the neural network.
Furthermore, the high accuracy for the reconstruction of synthetic data confirms that the chosen network architecture PIFu is adequate for the given task.
The reconstruction of experimental image data results in physically reasonable shapes of the gas-liquid interface for the complete droplet dynamics during impact with a high volumetric accuracy, as shown by the low uncertainty and bias errors of the reconstructed volume.
For the case with the most severe deformation of the droplet, that is impingement of the structured hydrophobic PDMS-substrate an uncertainty of $\sigma_V = 6.6\%$ and bias error of $\delta_V = 6.2\%$ of the reconstructed volume was found.
The reconstruction of droplet impingement on the structured PLA-substrate yielded lower errors with $\sigma_V = 5.7\%$ and $\delta_V = 4.9\%$, while the volumetric errors for the case with less severe and axis-symmetrical deformation were significantly lower, reaching $\sigma_V = 3.5\%$ and $\delta_V = 4.7\%$.
These results demonstrate the successful application of the neural network trained on synthetic data to the real world task and, consequently, validate the general approach of using synthetic data for the optimisation of the neural network.
Furthermore, this indicates that the synthetic data generation already reaches a sufficiently high quality, i.e. a good agreement with the real data distribution.

\subsection{Effectiveness of glare points}

The neural network trained on synthetic images that contain glare points from lateral light sources (\textit{DFS2023C}) consistently reached a higher quality reconstruction compared to the network that was trained on pure shadowgraph images (\textit{DFS2023E}). 
During the optimization of the network \textit{DFS2023C} already reaches a better training and validation performance, as indicated in Figure~\ref{fig:3D-IOU_train_val} by a larger $3D-IOU$, which signifies a better agreement of the reconstruction with the ground truth data.
The results for the reconstruction of images recorded in the experiments confirm this observation.
Overall, \textit{DFS2023C} reached lower uncertainties and bias errors for the reconstruction in all three tested cases of droplet impingement on a flat and the two structured surfaces, which is furher detailled in Tables~\ref{tab:uncertainties_all} and \ref{tab:bias_all}.

Furthermore, the reconstruction of \textit{DFS2023C} was more consistent in time, as indicated by a much lower frame-to-frame difference in the reconstruction in comparison to \textit{DFS2023E} (see Figure~\ref{fig:Contour_OOP_Rec}).
These results are further supported by the high frequency fluctuation in the reconstructed volume, that is evident for \textit{DFS2023E}, but not apparent for \textit{DFS2023C}, as can be seen in Figure~\ref{fig:vol_fdm_45}.
The employment of glare points by the network during reconstruction also furthers a more accurate local depth estimation, in particular for shapes that are different to the training data, as seen in Figure~\ref{fig:PDMS_depth_estim}.
The comparison of the in-plane contour for the reconstruction and the images recorded in the experiments, illustrated in Figures~\ref{fig:DFS2023C_PDMS_0_InP} and \ref{fig:DFS2023E_PDMS_0_InP}, reveals that both versions of the network perform a highly accurate reconstruction in the image plane.
Consequently, this leaves the depth estimation, related to the out-of-plane contour as the sole source for errors in the reconstruction. 

The higher error in the out-of-plane reconstruction was an expected result as the available features in the image are much more sparse in the case of images with glare points, or even missing in the case of a pure shadowgraph input in comparison to the in-plane reconstruction, for which the two-dimensional droplet contour is available.
There are two key findings that can be derived from this observation.
First, the neural network can be learned for a depth estimation that relies completely on the two-dimensional contour of a shadowgraph.
Second, the glare points successfully encode additional three-dimensional information that is considered by the neural network during reconstruction, which leads to a significant improvement in the depth estimation.
Thus, the presented results demonstrate the effectiveness of glare points for constraining the global and local reconstruction of the three-dimensional gas-liquid interface, which leads to a higher reconstruction accuracy.

\subsection{Reconstruction of obscured areas}

The results show that the neural network was able to reconstruct unseen regions of the gas-liquid interface in a physically meaningful manner.
This includes input images with self-occlusion, as seen in Figure~\ref{fig:rec_occlusion}, as well as the wetting state of structured surfaces, indicated by Figure~\ref{fig:PDMS_wetting} and finally, the learned depth estimation, as discussed previously.
These results suggest that the neural network is able to learn an approximation of the underlying droplet dynamics that is applied to fill in missing information in the experimental data in a physically reasonable way during reconstruction.
Consequently, the data-driven approach can be learned for a physically correct reconstruction by training data that accurately represents the underlying physics of the problem, such as the direct numerical simulation that was used in this study.
However, it should be noted that the quality of the reconstruction is lowered in the occluded regions due to artefacts and higher frequency errors of the reconstruction, which falls in line with previous reports \cite{saito2019}.

Furthermore, the reconstruction of the wetting state of structured surfaces was only successful if the surface structure was visible in the experiments at $\alpha=0^\circ$ observation angle, as seen in Figure~\ref{fig:PDMS_wetting}.
This effect also appeared for a version of the neural network that was exclusively trained on simulations of droplet impingement on structured surfaces (\textit{DS2022}) and thus was found to be independent from ambiguities due to training examples of the wetting of flat surfaces in the training dataset of \textit{DFS2023C}.
These results indicate that features in the image are prioritised by network over the learned knowledge from the training data and, more specifically strictly abided, as evident from the straight contour of liquid-solid contact area in the reconstruction (see e.g. Figure~\ref{fig:Contour_OOP_Rec}).

Moreover, the results highlight the relevance of pre-processing the input images with a binary mask.
The mask, which is composed of the shadowgraph contour of the droplet and the contour of the surface structure determined from the images recorded in the experiments prior to the impact of the droplet, as described in section~\ref{subsec:volumetric_reconstruction}, imposes this information on all later frames, where the counter of the droplet would conceal the liquid-solid boundary.
Thereby the masking furthers the reconstruction as only relevant information from the experiment is pre-selected and passed onto the neural network.
Consequently, the network does not need to learn to differentiate substrate from droplet, resulting in a simplified reconstruction.

However, care has to be taken for the detection of the substrate in the image, as errors from the position of the ground translate into an erroneous mask, which in turn introduces an error to the volumetric reconstruction that is related to the extent and shape of the droplet close to the contact area.
This source of error is a possible be a cause for the low frequency oscillation of the reconstructed integral volume observed in Figures~\ref{fig:vol_fdm_45} and \ref{fig:vol_pdms_90}.


\subsection{Versatility of the data-driven approach}

It is demonstrated that the neural network can accurately predict the gas-liquid interface during a separation of the droplet and subsequent coalescence during droplet impingement in the regime of partial rebound.
Furthermore, the small droplet that is ejected at high velocities from a Worthington jet is successfully reconstructed.
The ability to reconstruct these topological changes elucidates the advantage of the three-dimensional representation by a level-set method that underlies the PIFu algorithm.
Moreover, the discussed topological changes were not represented in the training dataset, which indicates a certain capability for an extrapolation beyond the training data cases in regards to fluid mechanical regimes.
It should be noted that this does not directly imply an extrapolation in terms of the training data distribution.
It can be concluded that the strict abidance of the network to the image features, in particular the contour of shadowgraph, allows for a reconstruction of unknown shapes of the gas-liquid-interface.
In this case the depth estimation appears to be supported by glare points, as can seen in Figure~\ref{fig:PDMS_depth_estim} and previously elaborated.

\subsection{Synthetic data generation}

As discussed in subsection~\ref{subsec:synthetic_data} a discrepancy of the synthetic training images to real images recorded by the experiments reduces the performance of the neural network and thus the applicability of synthetic training data \cite{Csurka2017,Shrivastava2017}.

The model for synthetic data generation does not consider all optical effects that are involved in the image creation in the experiment, in order to allow for an acceptable render time.
Therefore higher order glare points and reflection on the substrate are not modelled, but only the zeroth order glare points resulting from the lateral light sources and the first order glare point from the backlight.
Polarisation is not considered, since the LED lights are unpolarised.
The light sources are assumed to have a homogeneous distribution in brightness, which finds a good agreement to the experiments in which the light is homogenised through an optical diffuser.
The index of refraction is assumed to be constant, and therefore dispersion is not modelled.
However dispersion only needs to be considered for the $p=1$ glare point that is produced by the backlight, since the lateral glare points only undergo direct interface reflection ($p=0$).
Since the LED light have a narrow-banded spectrum the effect of dispersion is small and can therefore be neglected, see \cite{Dreisbach2023}.

Since the shadowgraph contour and the glare points originate from different depth coordinates, their defocus is dependent on the focal length and aperture of the camera equipment and therefore specific to the experimental setup.
Furthermore, the shape of the $p=1$ glare point is influenced by the shape of the aperture, as can be seen in Figure~\ref{fig:comp_render_real}.
These effects are modelled by explicitly reproducing the objective lens and aperture in the render setup.
However, the contour of the $p=1$ glare points appears to be too crisp in the rendering compared to real experiments, which indicates that some of the optical phenomena that were not modelled are relevant here.
This might include dispersion, absorption or light scattering by pollution in the water.
It should be noted that the render engine uses numerical approximations for the calculation of reflection and diffusion, which potentially influence the outcome of the $p=1$ glare point.
Furthermore, since the experimental setup is mirrored in the render setup the projection of the droplet to the camera is closely approximated, so that perspective errors can be neglected.

The successful reconstruction of images recorded in the experiment by a neural network trained on rendered synthetic data demonstrates that the synthetic data already closely matches the real data.
Further improvement of the reconstruction accuracy could be reached by domain adaption \cite{Csurka2017}, which is aimed to modify the already rendered synthetic data to move its feature distribution closer to that of the real images. 


\subsection{Performance characteristics}

The neural network was trained for eight epochs on $37.300$ training samples, with a batch size of $12$ and an initial learning rate of $0.001$ that is reduced by a factor of ten after epochs four and six.
The training duration amounted to $58$\,h on a Nvidia RTX A5000 graphics processing unit.
The network requires $20$\,s on average per time step for the volumetric reconstruction at an output resolution of $512^3$ grid nodes.
It should be noted that due to the implicit representation of the surface, the reconstruction can be performed at an arbitrary resolution, and consequently a speed-up is possible by lower output resolutions \cite{saito2019}.
Furthermore, processing the predicted grid nodes by the marching cubes algorithms \cite{Lorensen1987} is required in order to obtain a mesh of the gas-liquid interface and PIFu uses the octary tree structure \cite{Meagher1982} for a more efficient and thus faster inference.
The inference time is required proportionally for the following processes on average, $13.2$\,s for the prediction by the neural network, $1.6$\,s for marching cubes and $5.2$\,s for data handling.

\section{Conclusions}
\label{sec:conclusion}

The positive results for the spatio-temporal reconstruction of the gas-liquid interface of an impinging droplet from monocular experimental recordings demonstrate the success of the proposed approach based on neural networks and synthetic training data generation.
The employment of synthetic image rendering from the results of direct numerical simulation proved to be an effective method for the generation of suitable training data, while eliminating potential errors due to the inherent discrepancy of experimental and numerical results.
The single-camera setup required for the proposed method is both cost-effective and easy to calibrate and, therefore, accessible to wide field of applications.

It was shown that the glare points produced by additional lateral light sources are suited for the encoding of further information on the three-dimensional shape of the gas-liquid-interface in the image, which is successfully exploited by the neural network during reconstruction and results in a higher accuracy of the reconstruction.

In particular, color-coded glare points improve the depth estimation of the neural network and, thus, allow for the reconstruction of the complex non-axisymmetric shape of the gas-liquid-interface for droplet impingement on structured substrates from any arbitrary azimuth angle.
These findings recommend the proposed method for the reconstruction of the three-dimensional droplet dynamics for an impact on substrates for which the orientation angle is not known a-priori, which consequently allows for an efficient characterisation of these substrates.

It was demonstrated that the proposed method can reach a high accuracy on synthetic data, which establishes that the chosen neural network architecture, as well as the implicit three-dimensional representation through a level-set function is well suited for the given task.
Furthermore, by the reconstruction of experimental cases from different fluid mechanical regimes it was shown that the neural network was able to learn a versatile model of the involved two-phase flow phenomena, that even allows for topological changes.


The successful reconstruction of obscured regions in the input images indicates that the neural network leverages the learned knowledge of the droplet dynamics from training data that is based on numerical simulation to fill in missing information in a physically reasonable way.
It can be concluded that training the neural network with limited numerical data -- in our case two direct numerical simulations -- already learns the network for a physically correct reconstruction.
This highlights the advantage of the proposed data-driven method over conventional approaches, which furthermore can be improved by more data and expanded to a broader range of application by training on new regimes.
While the proposed framework of synthetic training data generation based on the results from numerical simulation shows promising results for monocular reconstruction, it can be adapted to multi-view reconstruction methods in order to further improve accuracy.
Due to the flexibility of the chosen neural network architecture an adaptation is straightforward.



Further applications for the proposed monocular reconstruction approach are experiments where a complete and continuous imaging of gas-liquid interface cannot be guarantied at all times.
The volumetric reconstruction of the gas-liquid interface for droplets impacting at an angle and droplets in cross-flows appears straight-forward, while splashing droplets and sprays appear possible, as indicated by the successful reconstruction of secondary droplets.
Furthermore, the reconstruction of gas bubbles in liquids is feasible, as glare points can be produced on their gas-liquid interfaces by the same experimental setup, as used in this work.
Preliminary theoretical work on the basis of the findings by Sax \citet{Sax2023} indicate the existence of a scattering angle at which only $p=0$ glare points exist for air bubbles in water, at $theta \approx 78.5^\circ$.
The general framework of the proposed approach can be adapted for other measurement techniques that are suitable to encode depth information of the gas-liquid interface, such as structured light techniques.

Currently, the temporal coherence of the dataset is not yet exploited, however, the additional information that the three-dimensional shape of the gas-liquid interface changes smoothly in time can be used to regulate the reconstruction and consequently allow for a higher accuracy.
For this purpose neural networks for image sequence processing that can exploit the temporal coherence of data structures could implemented in the neural network architecture.
For example, the backbone feature extractor of the employed PIFu architecture \cite{saito2019} could be expanded with convolutional long short-term memory networks \cite{Donahue2015} or spatio-temporal transformer networks \cite{Zeng2020}.

Further prospects are offered by the direct introduction of physics to the neural network optimisation through physics-informed neural networks (PINN) \cite{Raissi2019}.
In this framework a neural network is trained to respect the underlying differential equations of the considered fluid mechanical problem, such as the Navier-Stokes \cite{Cai2021,Buhendwa2021}, Cahn-Hilliard or Allen-Cahn \cite{Zhao2021} equations, which allows for a prediction of fluid dynamical quantities with very little or even no data at all.
Previous works have already demonstrated the capability of PINNs for the reconstruction of three-dimensional velocity and pressure fields \cite{Cai2021} and two-phase flows \cite{Buhendwa2021}.


\section*{Data availability statement}
All data that support the findings of this study, including the trained neural networks and any supplementary files are available upon request.

\section*{References}
\bibliographystyle{iopart-num}
\bibliography{literature.bib}   

\newpage
\appendix
\section{Failure cases}
\setcounter{figure}{0}

\begin{figure}[htbp!]
\centering
    \includegraphics[width=1.0\columnwidth]{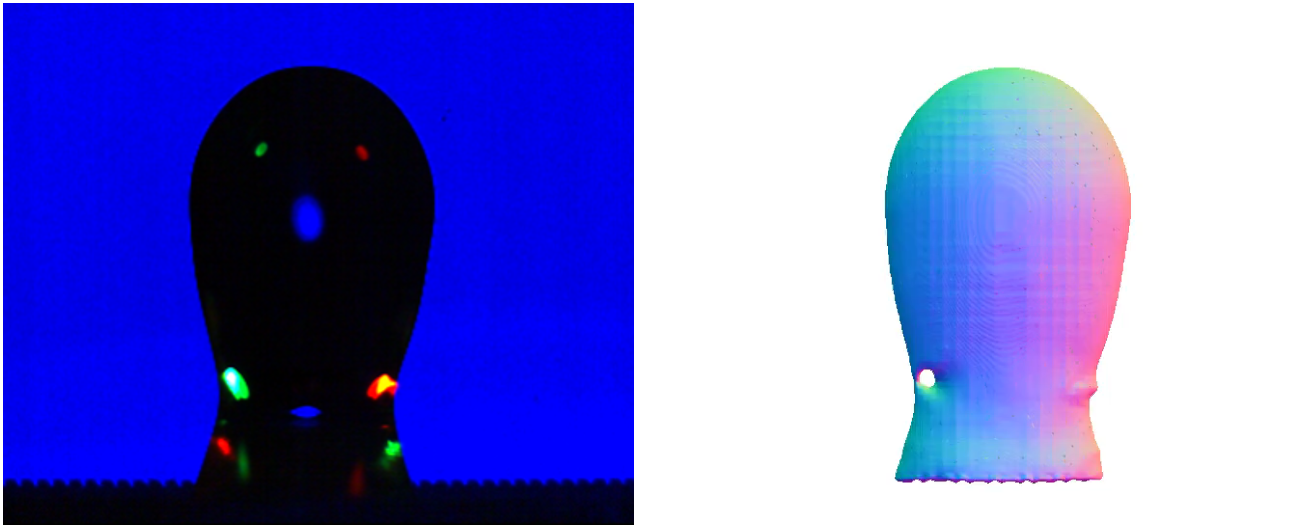}
    \caption{Input image frame with overexposure and higher order glare points (left) and the respective reconstruction (right).}
    \label{fig:PDMS_hard_case}
\end{figure}

\begin{figure}[htbp!]
\centering
    \includegraphics[width=1.0\columnwidth]{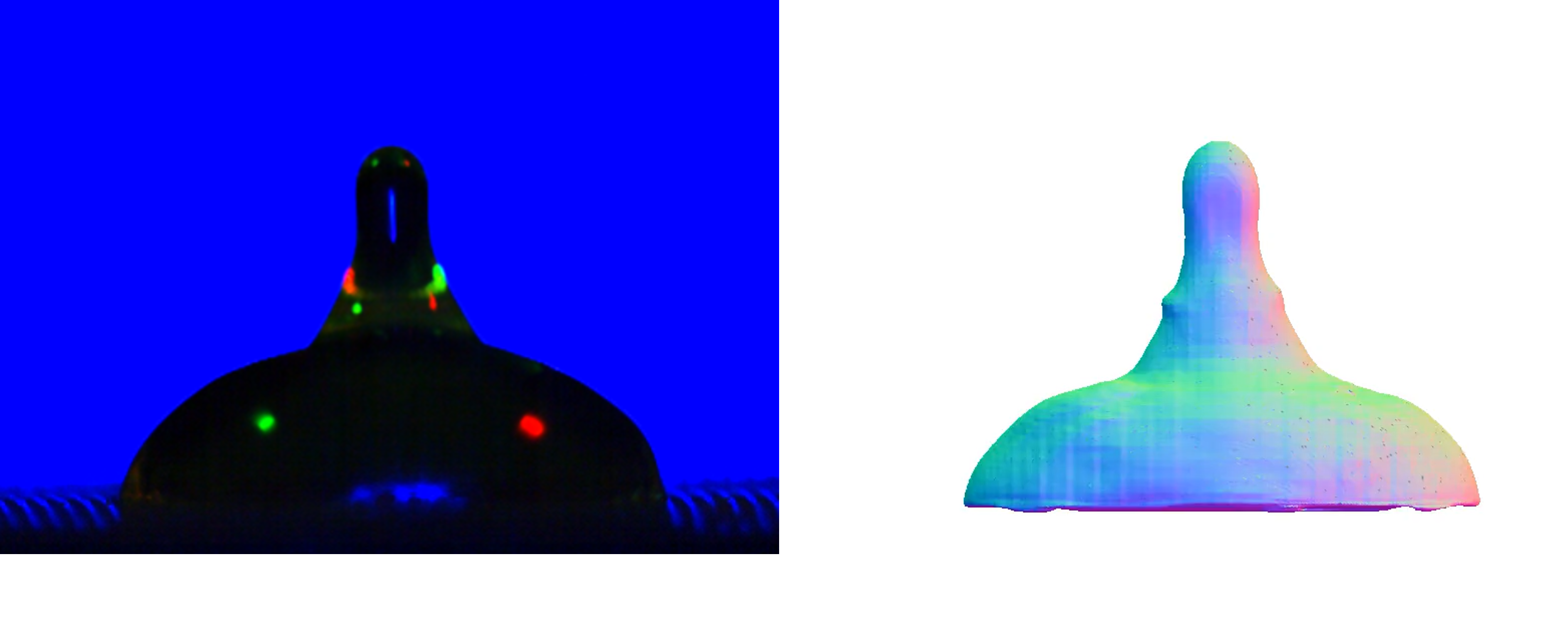}
    \hfill
    \includegraphics[width=1.0\columnwidth]{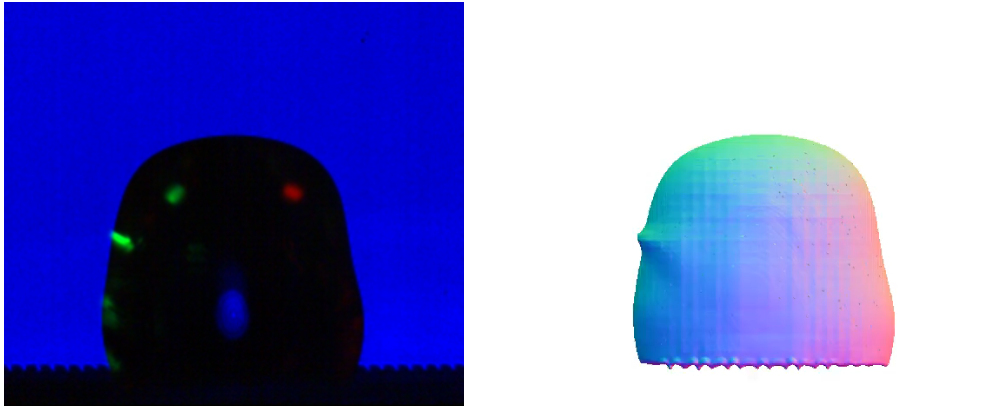}
    \hfill
    \includegraphics[width=1.0\columnwidth]{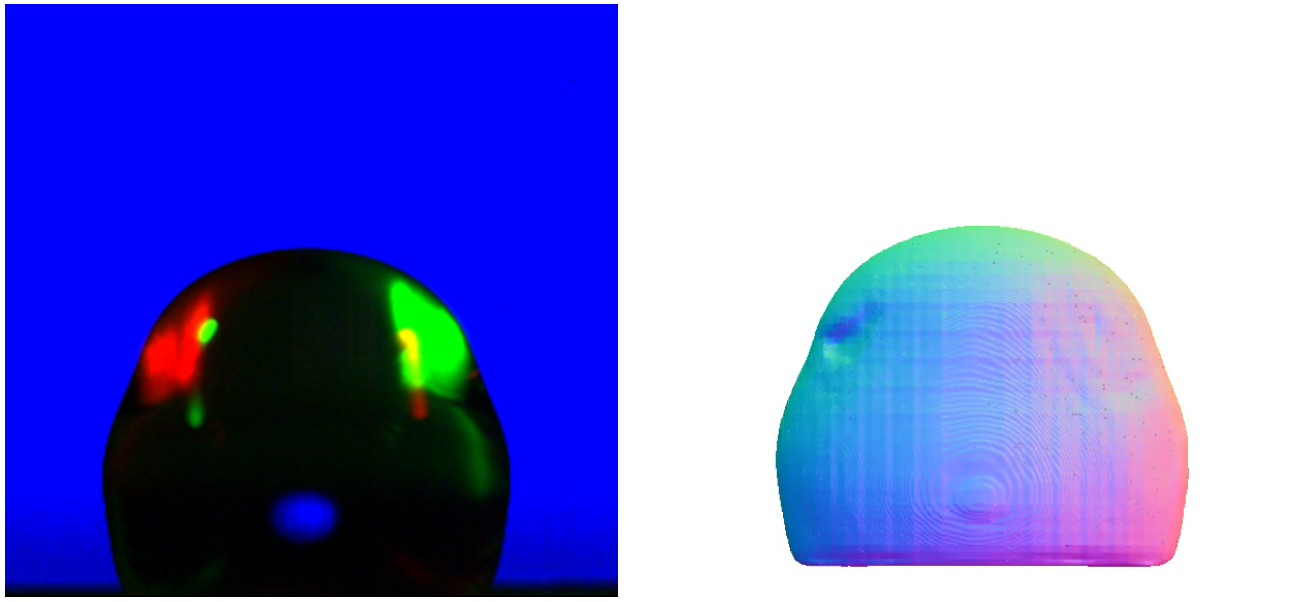}
    \caption{Input image frames with overexposure and higher order glare points (left) and the respective reconstructed volumes (right).}
    \label{fig:PDMS_hard_cases}
\end{figure}

The comparison of the images recorded in the experiments with droplet impingement on the structured PLA and PDMS-substrates reveals that a reflection of the incident light from the lateral illumination on the substrate occurs predominantly for the PDMS-substrate.
The lateral illumination has an $45^\circ$ incidence angle and therefore light can be reflected at the liquid-solid contact area after passing through the droplet and subsequently produce additional glare points on the gas-liquid-interface while exiting the droplet again (see Figure~\ref{fig:PDMS_hard_cases}, third example).
Additionally, the light can reflect on the substrate outside of the droplet and subsequently enter the droplet an an $-45^\circ$ incidence angle. 
Both mechanisms of reflection were confirmed to occur in the experiments.
The PLA-substrate was produced from a low reflective material in order to suppress spurious glare points from reflection and allow for an evaluation of the their influence on the quality of the reconstruction.
Furthermore, impingement on the hydrophobic PDMS-substrate lead to a higher degree of droplet deformation in comparison to the PLA-substrate.
The resulting complex shapes of the gas-liquid interface allowed for the emergence of higher order glare points \cite{Hulst1981, Hulst91} that are created by internal reflection of the light on the gas-liquid interface.
The consequence of the internal reflection from higher order glare point and reflection on the substrate within the droplet is that the glare points change their position to the other hemisphere of the droplet, as can be seen for the lowest pair of glare points in Figure~\ref{fig:PDMS_hard_case}. 

A second consequence of the internal reflections is the focusing of the incident light by the curved droplet contour, which can result in a magnification of the light intensity for higher order glare points, as obvious from Figure~\ref{fig:PDMS_hard_case} by the second row of glare points.
Overexposure in the images leads to color clipping and a loss of information in the experiments, as the clipped image channel cannot register any further increase in intensity \cite{Novak1990}.
As evident from the reconstruction result in Figure~\ref{fig:PDMS_hard_case}, color clipping in the regions of the bright glare points causes nonphysical artefacts in the reconstructed gas-liquid interface, such as the hole on the left and the dimple on the right of the reconstructed geometry.
Further examples of images with reflection and higher order glare points and their respective volumetric reconstruction can be found in Appendix Figure~\ref{fig:PDMS_hard_cases}.

The comparison of the uncertainty and bias errors of the reconstructed integral volume between droplet impingement on the PLA- and PDMS-substrate, as detailed in Tables~\ref{tab:uncertainties_all} and \ref{tab:bias_all}, reveals significantly lower errors for the reconstruction of the PLA case.
Furthermore the specific inspection of frames associated with a particular high volumetric error as indicated by Figure~\ref{fig:vol_pdms_90} point to an association with images that are affected by a higher degree of disturbance through reflections and overexposure.
These results indicate that reflection and higher order GP lead to higher errors in the reconstruction of the gas-liquid interface.
This effect has to be expected since the input images to the neural network have to be similar to its training data in order to reach a high quality in the reconstruction, as described in subsection~\ref{subsec:synthetic_data} and reflection on the substrate, as well as high order glare points were not modelled for synthetic training data generation.
Therefore, the unexpected input data causes a high uncertainty in the reconstruction and can even lead to the erroneous prediction of the local gas-liquid interface, as seen in Figure~\ref{fig:PDMS_hard_case}.

Conversely, these results further underline that the neural network considers the glare points during reconstruction in order to guide the reconstruction. 
Furthermore, the results indicate a certain robustness towards disturbance outside of local errors, as glare in unexpected regions (see Figure~\ref{fig:PDMS_hard_case}) and unexpectedly large glare points (Figure~\ref{fig:PDMS_hard_cases} third example) appear to be mostly disregarded by the neural network for the reconstruction of the global shape.

\newpage
\section{~}

\begin{figure}[htbp!]
\centering
    \includegraphics[width=1.0\columnwidth]{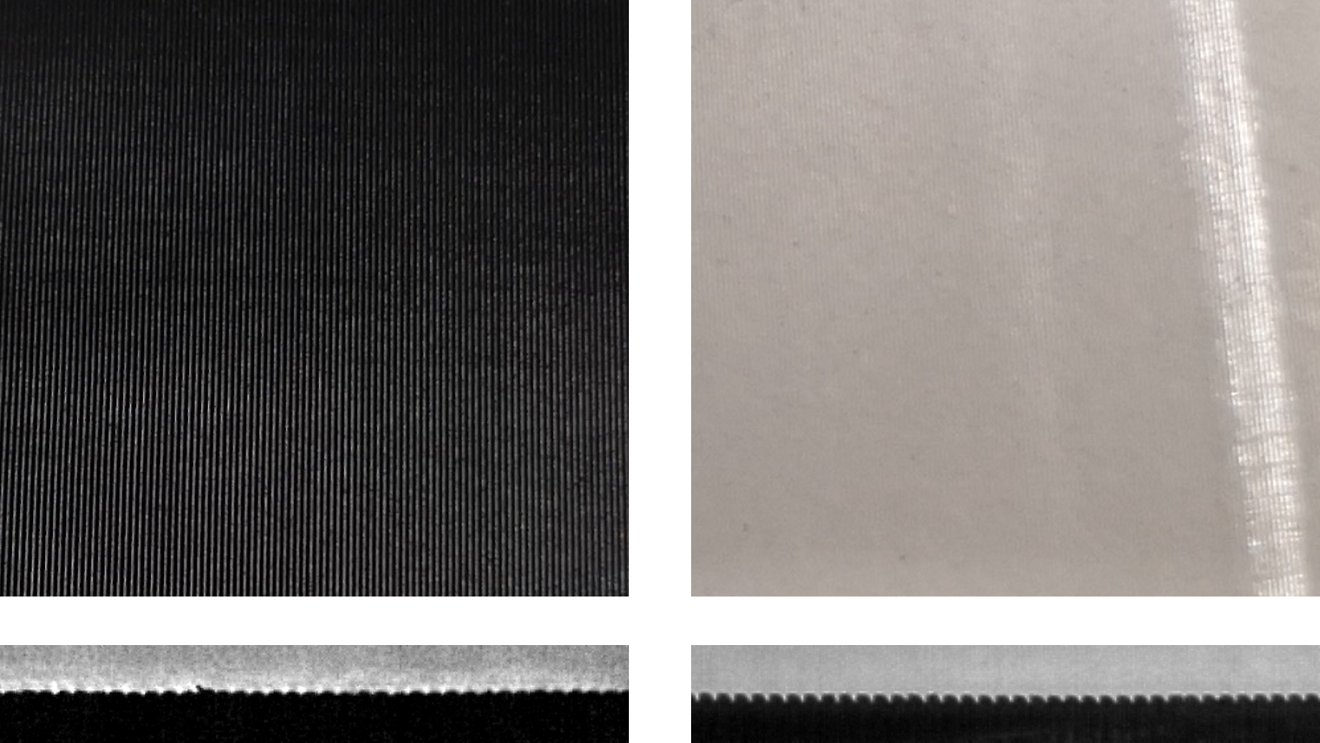}
    \caption{Photographs of the substrates with structured surface, 3D-printed polylactide (left) and Polydimethylsiloxane (right). The detail (below) shows the side view on the substrate at $0^\circ$ observation angle.}
    \label{fig:substrates}
\end{figure}

\begin{figure}[ht]
    \centering
    \includegraphics[width=0.85\columnwidth]{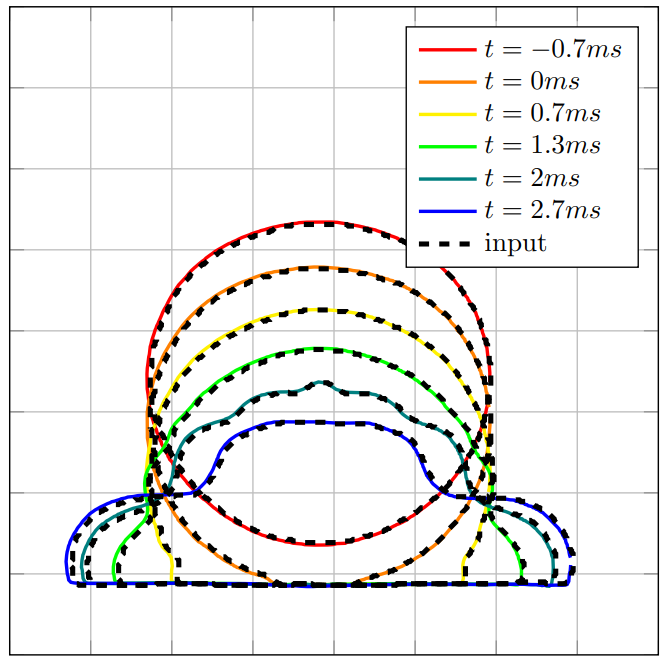}
    \caption{In-plane contour of the reconstructed droplet shapes over time (colored) in comparison to contour of the input shadowgraph (dashed black) for the reconstruction of droplet impingement on the structured PLA-substrate at $\alpha=90^\circ$ by \textit{DFS2023E}.}
    \label{fig:DFS2023E_PLA_90_InP}
\end{figure}

\begin{table}[ht]
    \centering
    \caption{Contact angles of the PLA and PDMS-substrates in parallel and transversal direction.}
	\begin{tabular}{l|rrrrrrr}		
		case & {$\theta_a$} & {$\theta_r$} & {$\theta_{eq}$} & {$\Delta \theta$} & {no. exp.}\\ \hline \rule{0pt}{1.0\normalbaselineskip}
            PDMS $0^\circ$  & 115 & 88 & 107 & 27 & 5 \\
            PDMS $90^\circ$ & 107 & 74 & 97 & 33 & 4 \\
            PLA $0^\circ$  & 113 & 59 & 76 & 54 & 2 \\
            PLA $90^\circ$ & 101 & 52 & 63 & 49 & 2 \\
	\end{tabular} 
	\label{tab:dyn_contact_angles}
\end{table}

\begin{table}[ht]
    \centering
    \caption{Uncertainty $\sigma_V$ of the reconstructed integral volume for \textit{DFS2023C} and \textit{DFS2023E} in percent of ground truth volume.}
	\begin{tabular}{l|rrrrrrr}		
		case & {DFS2023C} & {DFS2023E} \\ \hline \rule{0pt}{1.0\normalbaselineskip}
            flat            & 3.5 & 5.1 \\
            PLA $0^\circ$  & 7.0 & 7.9 \\
            PLA $45^\circ$ & 2.9 & 3.1 \\
            PLA $90^\circ$ & 7.2 & 6.8 \\
            PDMS $0^\circ$  & 9.3 & 9.5 \\
            PDMS $45^\circ$ & 3.9 & 2.9 \\
            PDMS $90^\circ$ & 3.7 & 4.3 \\
	\end{tabular} 
	\label{tab:uncertainties_all}
\end{table}

\begin{table}[ht]
    \centering
    \caption{Bias error $\delta_V$ of the reconstructed integral volume for \textit{DFS2023C} and \textit{DFS2023E} in percent of ground truth volume.}
	\begin{tabular}{l|rrrrrrr}		
		case & {DFS2023C} & {DFS2023E} \\ \hline \rule{0pt}{1.0\normalbaselineskip}
            flat            & 4.7 & 6.2 \\
            PLA $0^\circ$  & 9.9 & 11.7 \\
            PLA $45^\circ$ & 0.3 & 6.3 \\
            PLA $90^\circ$ & 4.5 & 6.1 \\
            PDMS $0^\circ$  & 5.1 & 6.9 \\
            PDMS $45^\circ$ & 7.8 & 9.3 \\
            PDMS $90^\circ$ & 7.2 & 10.0 \\
	\end{tabular} 
	\label{tab:bias_all}
\end{table}

\end{document}